\documentclass[a4paper,11pt]{article}
\pdfoutput=1 

\usepackage{jheppub} 

\usepackage[english]{babel}
\usepackage[utf8x]{inputenc}
\usepackage[T1]{fontenc}

\usepackage{amsmath}
\usepackage{graphicx}
\usepackage{tikz-feynman}
\usepackage{float}
\usepackage{subfigure}
\usepackage{url}
\usepackage[colorinlistoftodos]{todonotes}
\usepackage{comment,color}

\title{\boldmath Final bound-state formation effect on dark matter annihilation}

\author[a]{Xinyu Wang}
\author[a]{Fucheng Zhong}
\author[a]{Feng Luo}

\affiliation[a]{School of Physics and Astronomy, Sun Yat-sen University, Zhuhai Campus, 2 Daxue Road, Xiangzhou District, Zhuhai, P. R. China}

\emailAdd{wangxy525@mail2.sysu.edu.cn}
\emailAdd{zhongfch@mail2.sysu.edu.cn}
\emailAdd{luofeng5@mail.sysu.edu.cn}

\abstract{If two annihilation products of dark matter (DM) particles are non-relativistic and couple to a light force mediator, their plane wave functions are modified due to multiple exchanges of the force mediator. 
This gives rise to the final state Sommerfeld (FSS) effect. 
It is also possible that the final state particles form a bound state. 
Both the FSS effect and final state bound-state (FBS) effect need to be considered in the calculation of the DM relic abundance. 
The annihilation products can be non-relativistic if their masses are comparable to the annihilating DM particles. 
We study the FSS and FBS effects in the mass-degenerate region using two specific models. Both models serve to illustrate different partial-wave contributions in the calculations of the FSS and FBS effects.
We find that the FBS effect can be comparable to the FSS effect when the annihilation products couple strongly with a light force mediator.
Those effects significantly modify the DM relic abundance.}

\keywords{dark matter --- bound state formation --- final state sommerfeld effect --- relic abundance}

\begin{document} 
\maketitle
\flushbottom

\section{Introduction}
\label{sec:intro}

The standard cosmological model ($\Lambda$CDM) successfully described the large scale structure evolution of our universe. The main constitute of matter component in $\Lambda$CDM, called dark matter (DM), which still not yet be detected, is one of the most outstanding puzzles in contemporary physics. The cosmology observation accuracy about DM abundance has reached percent level\cite{Planck:2018vyg}. The relic abundance of DM needs to be considered beyond the perturbation calculation. 

Weakly Interacting Massive Particle (WIMP)\cite{Jungman:1995df} is one of the most popular dark matter candidates. In the standard paradigm, the relic abundance of WIMP dark matter is usually given by the thermal freeze-out mechanism\cite{PhysRevLett.39.165,Kolb:1990vq}. If there exists long-range force between two non-relativistic moving particles, non-perturbation effect needs to be considered. This effect can be calculated by ladder diagram in Quantum Field Theory\cite{Petraki:2015hla}. It can also be approximately calculated by quantum mechanics that considering the two particle pair wave function is modified by the long-range force. 
The DM particles are non-relativistic during freeze-out in the most models about WIMP (so we called it cold dark matter). Many works focus on the non-perturbation effects of annihilation particles, for example, the Sommerfeld effect and bound state effect\footnote{The Sommerfeld effect is caused by the scattering-state wave function (continuous spectrum), the bound state is formed due to bound-state wave function (discrete spectrum).}.

Previous works about the bound state effect have been focusing on the initial DM or coannihilator
pairs\cite{https://doi.org/10.1002/andp.19314030302,Feng:2009mn,vonHarling:2014kha,Ellis:2015vaa,Liew:2016hqo,Harz:2019rro,Harz:2018csl,Fukuda:2018ufg,Smirnov:2019ngs,Petraki:2015hla,Becker:2022iso,Biondini:2018ovz,Biondini:2018xor,Biondini:2021ycj}. However, when annihilation products have a coupling with a light force mediator, and have been mass-degenerate with the initial DM, those non-perturbation effects also occur in final state particles. In the early universe, the initial annihilating particles energy obey  Maxwell-Boltzmann distribution. So they can have enough energy to annihilate into heavier particles (Forbidden DM \cite{DAgnolo:2015ujb} or Impeded DM \cite{Kopp:2016yji}). 
Recently, the final state Sommerfeld effect (FSS) has been considered in the DM relic abundance calculation\cite{Cui:2020ppc}. It is showed that the $s$-wave FSS has a significant influence on the DM relic abundance. We naturally extend to $p$-wave FSS effect in this work.

In fact, the FBS formation has been discussed routinely in Standard Model (SM), including $e^+e^-\rightarrow(\mu^+\mu^-)$ \cite{PhysRevLett.35.1605}, $q\bar{q}\rightarrow(\mu^+\mu^-)g$ \cite{Chen:2012ci}, where the bound state are formed by exchanging the SM gauge bosons. 

Usually, the leading FBS formation should be the $2 \to 1$ process (two dark matter particles annihilate into FBS without emission). But the sub-leading process (two dark matter particles annihilate into FBS with emission) may be not negligible. Because the leading process can not always happens when the incoming particles energy is larger than the mass of the FBS. Another reason is, the sub-leading process can form a $s$-wave FBS, while the leading process can merely form a $p$-wave bound state. We provide Model II to illustrate this situation in Section \ref{sec:Model II}.

FBS can arise due to non-confining forces (hydrogen and positronium) or confining forces (hadronic bound states) \cite{Hoyer:2014gna}. Another class of bound states – non-topological solitons, has also been considered in the context of DM\cite{Kusenko:2001vu,Kusenko:2004yw}. Here we consider FBS formed due to non-confining interactions, and calculate the cross sections for FBS formation in the non-relativistic regime, which is relevant for cosmology and DM indirect detection signals. If the FBS can exist as a portal between DM sector and SM particles, it has the possibility to provide new detectable signals. For instance, the FBS has different energy levels like hydrogen, its decay and transition give the spectrum of SM particles. For example, the FBS formed due to exchange dark photon, which has a kinetic mixing with photon\cite{Baldes:2020hwx,Curtin:2014cca}.




The rest of our paper is organized as follows. In Section \ref{sec:FBS}, we introduce the FBS formation conditions and assumptions, also exhibit the physical picture of FBS. In Section \ref{sec:Thermal averag} and \ref{sec:Boltzmann equation}, we show the paradigm of calculating the DM relic abundance with FBS and FSS effect in the thermal freeze-out scenario. Section \ref{sec:Model I} is the analytical calculation about the cross sections for DM model I, and the numerical results about thermal-averanged cross sections, FBS effect on DM relic abundance and brief comment on the results. Section \ref{sec:Model II} is the same like Section \ref{sec:Model I} but for DM model II. DM model II is proposed to demonstrate the difference from Model I when considering the conservation of the angular momentum. The Models, both show the FBS can have important effect on relic abundance. Finally, in Section \ref{sec:comclusion} we summarize our conclusions.


\section{Final bound-state formation effect}
\label{sec:FBS}

\subsection{Final bound-state formation effect}
\begin{figure}[H]
  \centering
  \subfigure{
  \includegraphics[width=0.45\textwidth]{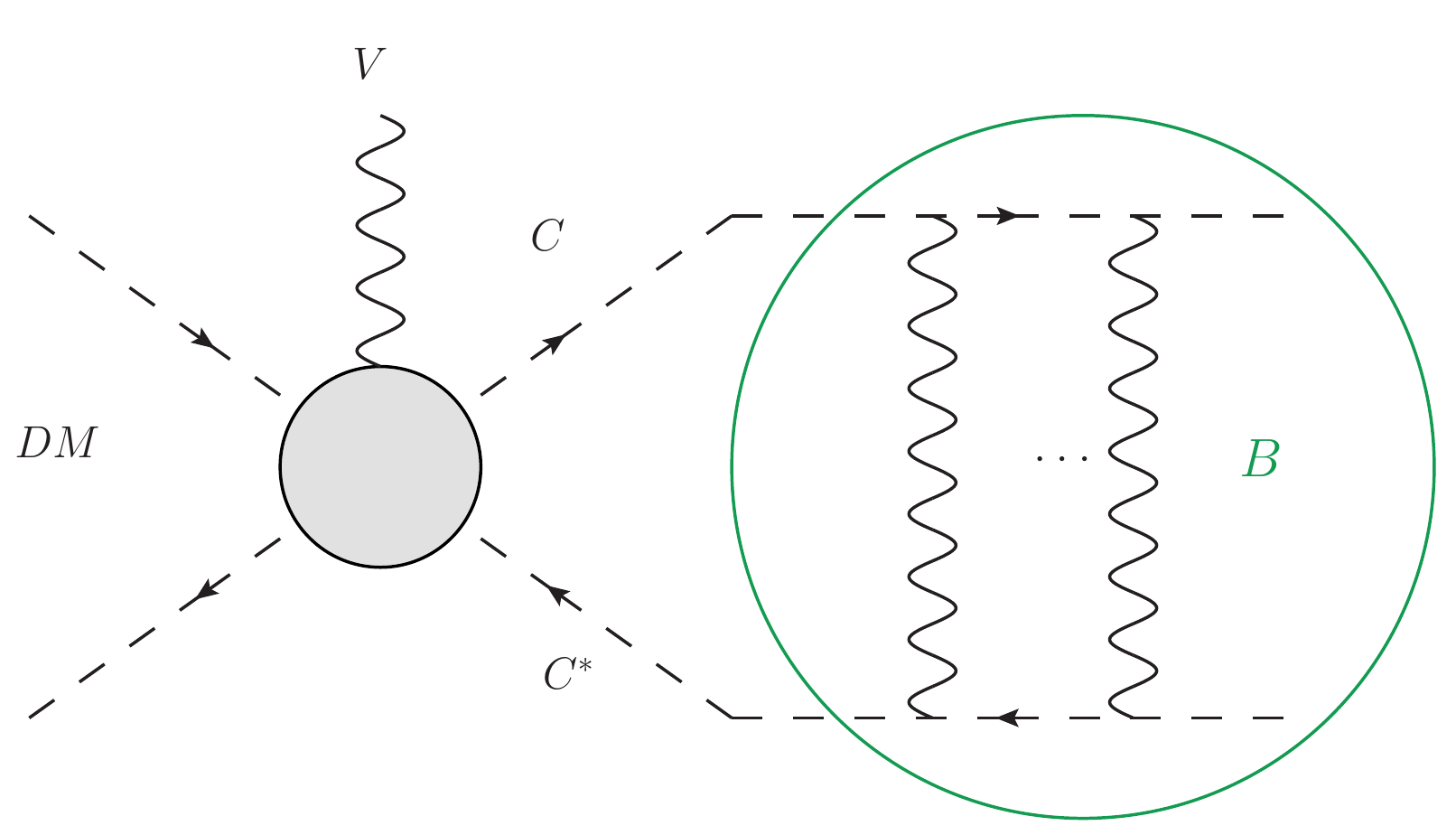}}
  \hspace{0in}
  \subfigure{
  \includegraphics[width=0.45\textwidth]{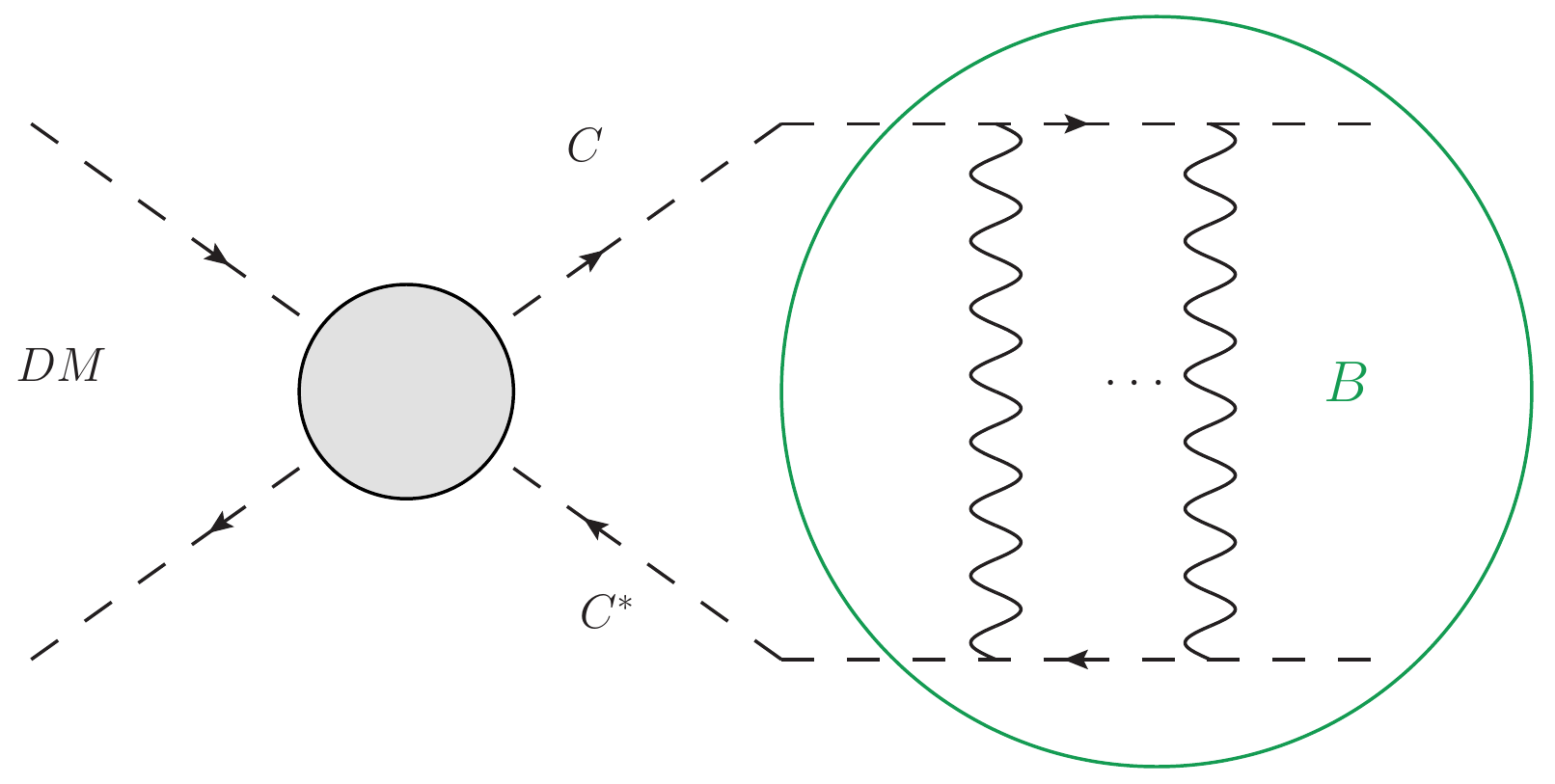}}
  \caption{\it Feynman diagram for the FBS formation. The two pictures both describe two incoming DM particles annihilate into a FBS, which is formed by $CC^*$ pair. The difference between the two panel: in the left panel, the incoming DM energy exceeds the FBS mass, the extra energy is carried away by a vector boson; in the right panel, the incoming DM energy equals the FBS mass.}
    \label{fig:BSF}
\end{figure}

When the annihilation products are non-relativistic and have a coupling with a light force mediator, bound state can form between final products. For simplicity, we consider two dark matter particles annihilate into two final state particles, the two final particles will form a bound state, the excessive energy of incoming particles is carried away by emitting a vector boson. This process can be described by Fig. \ref{fig:BSF}. 

It is not hard to understand the FBS formation with the analogy of positronium or true muonium formation (\cite{Karshenboim:2003vs,PhysRevLett.35.1605,CidVidal:2019qub}). The FBS has different energy levels like positronium, the most contributive channel is the $s$-wave FBS formation, and then the $p$-wave FBS formation. Considering the selection rule, in some situations, the $s$-wave channel can not happen, it can only start from $p$-wave channel. We propose Moldel II in Section \ref{sec:Model II} to demonstrate the situation in which it can only start from $p$-wave channel.

Another point is about the FBS life time. In this work, the FBS formed due to the non-relativistic final state particles exchange Abelian light mediator, just like the positronium and true muonium. They are not stable state and finally dissociate or decay, but it does not matter in our calculation. The final particles are regarded as a portal between DM sector and SM, they finally decay into SM particle and fade out as universe cools down. However, when DM freeze-out, they are still deemed to be within the thermal bath of SM particles. We assume the annihilation products have strong enough couplings, directly or indirectly, with SM particles, so that $C$ and the SM particles can maintain kinetic equilibrium and have the same temperature during the dark matter freeze-out. If $C$ couples with the SM particles via a dark photon $V$, the coupling $\alpha_{V}$ we considered is large enough ($\alpha_{V}=0.02$ is already large enough) for this assumption to be valid \cite{PhysRevD.82.083525,vonHarling:2014kha}. The final state particle $C$ (and $C^*$) may decay into other lighter particles which carry the same $U(1)$ charge. Therefore, the current experimental bounds on the dark photon parameters may not be directly translated to the bounds on $\alpha_{V}$ and $m_C$. On the other hand, it is possible to let the $CC^*$ particles annihilation cross section be large enough so that $C$ and $C^*$ can maintain thermal equilibrium during DM freeze-out. We only use the assumption that the final state particles are in thermal equilibrium during DM freeze-out, and building a full model is beyond the scope of this work. So we can take the simple Maxwell-Boltzmann distribution for them when we calculate the DM relic abundance. FBS exists as a portal, also  brings many interesting DM indirect detect signals (\cite{Baldes:2020hwx,Curtin:2014cca,Slatyer:2021qgc}), but we will not discuss here. Next we briefly explain why the existence of FBS  will influence the DM relic abundance. 

First, during freeze-out, most of the DM particles are moving with non-relativistic velocities. As the annihilation products of DM, the final state particles are naturally non-relativistic when their mass are degenerate with DM. As mentioned above, if the final state particles can exchange light vector boson, or in non-relativistic approximation, exits long range force, and the revolution time is smaller than the life time of the FBS components \cite{1998EPJC....2..345F}, it can form FBS.

Second, in the early universe, Standard Model particles, DM particles coupling with SM, are all in a thermal bath. The velocity of the particles in the thermal bath can be approximatively described by Maxwell-Boltzmann distribution. Therefore, there are always DM with enough energy that can annihilate into  heavier particles\cite{Griest:1990kh,DAgnolo:2015ujb,Kopp:2016yji}. In this case, the FBS formation processes are permitted even if the final state particles have larger mass than the DM. The ``forbidden'' cases are considered more significant because the FBS formation without emission can happen. In such consideration, we naturally study the FBS effect on the DM relic abundance.




\subsection{Thermal average including the ``forbidden'' case} 
\label{sec:Thermal averag} 
To calculate the effect of FBS or FSS effects on the DM relic abundance, we need to average the cross sections over the momentum distribution of DM in the early universe. All the initial and final particles are in the plasma, the Maxwell-Boltzmann distribution is parameterized by the energy of DM particles
\begin{equation}
\label{eq:Max-Bol eq}
  f(E)\propto e^{-\frac{Ex}{m_D}}
\end{equation} 
where $x\equiv m_D/T$, $m_D$ is the DM mass, $E$ is the energy of DM particle. The thermal-averaged cross section times relative velocity of DM annihilation are given by
\begin{equation}
  \langle \sigma v\rangle =\frac{\int\sigma ve^{-\frac{E_1x}{m_D}}e^{-\frac{E_2x}{m_D}}d^3\mathbf{p}_1d^3\mathbf{p}_2}{\int e^{-\frac{E_1x}{m_D}}e^{-\frac{E_2x}{m_D}}d^3\mathbf{p}_1d^3\mathbf{p}_2}.
\end{equation}

By changing the integration variables, the thermal-averaged cross section finally can be expressed as \cite{1991NuPhB.360..145G}
\begin{equation}
\label{average we use}
    \langle \sigma v\rangle=\frac{1}{8m_D^4TK_2^2(m_D/T)}\int_{s_{min}}^\infty\sigma(s-4m_D^2)\sqrt{s}K_1(\sqrt{s}/T)ds,
\end{equation}
where $s=(p_1+p_2)^2$ is the Mandelstam variable and $K_i$ is modified Bessel functions of order $i$. The integral must be from $s_{min}$ rather than $4m_D^2$ to take into account the threshold mentioned above.

\subsection{DM relic abundance}
\label{sec:Boltzmann equation}
In this section we discuss the relic abundance calculation including FBS formation. We assume a simple condition that the annihilation products quickly thermalize and their number densities equal the thermal equilibrium values as we mentioned in Section \ref{sec:FBS}. Therefore, only $one$ Boltzmann equation is needed to calculate the DM $yield$, which is the ratio of the DM density to the entropy density, $Y\equiv n/s$. The Boltzmann equation can be written as
\begin{equation}
    \label{eq:Boltzmann equation}
       \frac{dY_D}{dx}=-\frac{xs}{H(m_D)}\left(1+\frac{T}{3g_{*s}}\frac{dg_{*s}}{dT}\right)\langle \sigma v\rangle(Y_D^2-Y_{Deq}^2).
\end{equation}
Noting that:
\begin{equation}
    \begin{aligned}
        n_{eq}=\frac{T}{2\pi^2}gm_D^2K_2(x),\ s=\frac{2\pi^2}{45}g_{*s}m_D^3/x^3, \ H(m_D)=\left(\frac{4\pi^3G_Ng_*}{45}\right)^{\frac{1}{2}}m_D^2
    \end{aligned}
\end{equation}
where $g$ is the DM degrees of freedom; $K_2(x)$ is the modified Bessel function of the second kind; $G_N$ is the gravitational constant; $H(T)$ is the Hubble parameter; $g_{*s}$ and $g_*$ are the numbers of effectively massless degrees of freedom associated with the entropy density and the energy density, respectively. 

The total thermal-averaged cross section of different channels/effects is
\begin{equation}
    \langle\sigma v\rangle_{all}=\langle\sigma v\rangle_{FSS}+\langle\sigma v \rangle_{B}+\langle\sigma v \rangle_{BV},
\end{equation}
where $\langle\sigma v\rangle_{FSS}$, $\langle\sigma v \rangle_{B}$ and $\langle\sigma v \rangle_{BV}$ are the thermal-averaged cross sections for FSS-corrected annihilation, FBS formation without and with boson emission, respectively.
We just need to pump the total thermal cross section $\langle\sigma v\rangle_{all}$ into Boltzmann equation \ref{eq:Boltzmann equation}, than solve the $yield$ with FSS and FBS formation effect. The DM relic abundance is given by
\begin{equation}
    \Omega_D h^2= 2.755 \times 10^8  \frac{m_D}{GeV}Y_{D,0},
\end{equation}
where $h$ is the present-day dimensionless Hubble parameter; $Y_{D,0}$ is the DM $yield$ solved by Boltzmann equation and take the value at the limit $x \to \infty$.

\section{Model I}
\label{sec:Model I}

In this section, we employ a simple scalar QED-like model, which carry the light force mediator, to illustrate the FBS formation effect and FSS effect on relic abundance. We will calculate the cross section of FBS formation, and the cross section of the channel with FSS effect according to the model, then follow the process in Section \ref{sec:FBS}, and give the numerical results.

In this model, DM consisting of a real scalar particle $D$, coupled to a complex scalar particle $C$ via a four-point interaction. As DM annihilation products, $C$ also has QED-like couplings with a light real vector boson $V$ (for example, a dark photon). The model is summarised by the Lagrangian: 
\begin{equation}
    \mathcal{L}_I\supset|D_\mu C|^2+\frac{1}{2}g_DD^2|C|^2,
\end{equation} 
where $D_\mu=\partial_\mu-ig_VV_\mu$ is the covariant derivative. The coupling constant of the four-point interaction is $\mathnormal{g}_D$. The $V_\mu$ stands for light real vector boson. 

In $zero$ temperature, the vector boson mass is $zero$. While, in the thermal bath, the Coulomb force gets screened by the thermal plasma, this can be described by a vector Debye mass \cite{Cirelli:2007xd,RevModPhys.53.43}:
\begin{equation}
    \label{eq:Debye mass}
    m_V\sim  g_VT,
\end{equation}
where $T$ is the thermal bath temperature. The incoming DM particles $D$ with mass $m_D$, the outgoing particles $C$ with mass $m_C$. We assume that the vector boson mass all comes from the Debye mass. However, in calculating the FSS and FBS effects, we still use a Coulomb-like potential, since DM freeze-out happens with a temperature much smaller than the DM mass (typically $T_{freeze-out}\thicksim m_D/25$). The Debye mass is scanning when temperature decreases, so the Sommerfeld factor $S_f$ has resonant behavior \cite{PhysRevD.82.083525}. However, the difference between the thermal-averaged FSS-corrected cross sections is within several percent when we use the Coulomb-like potential and the Hulthen potential (which is a good approximation for the Yukawa potential). Therefore, during and after freeze-out, a Coulomb-like potential is a good approximation.

\subsection{Cross sections for Model I}
\label{Cross sections for Model I}
\subsubsection{Direct annihilation with FSS effect}

In this model DM particle $D$ can directly annihilate into $C$ via four- point interaction. 
\begin{figure}[H]
  \centering
  \subfigure{
  \includegraphics[width=0.3\textwidth]{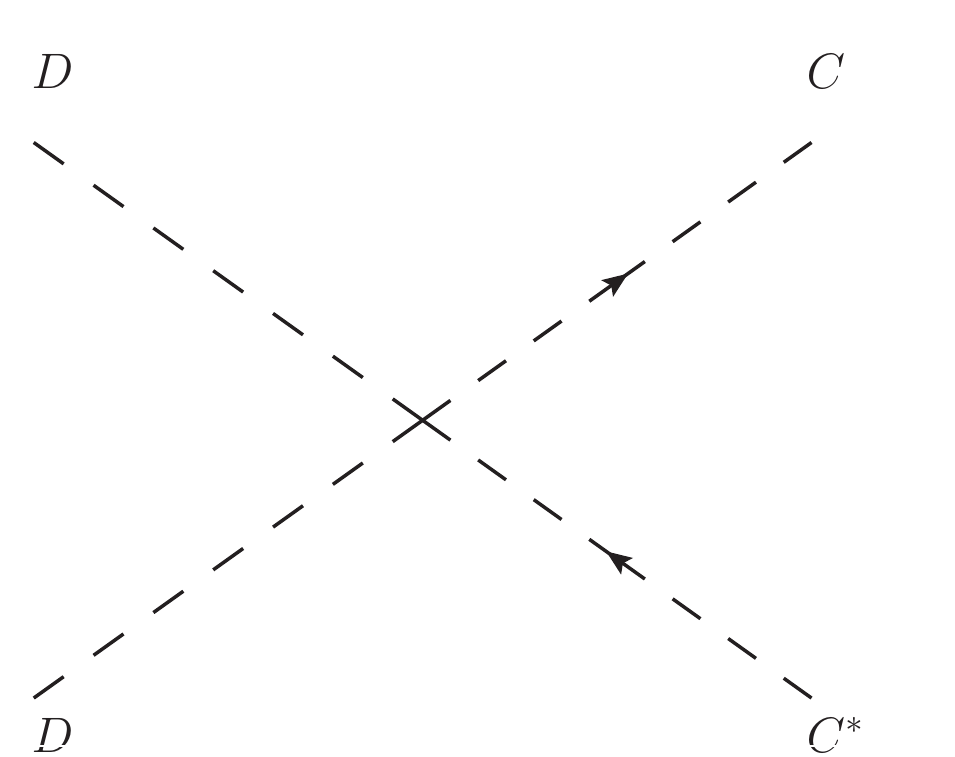}}
  \hspace{0in}
  \subfigure{
  \includegraphics[width=0.5\textwidth]{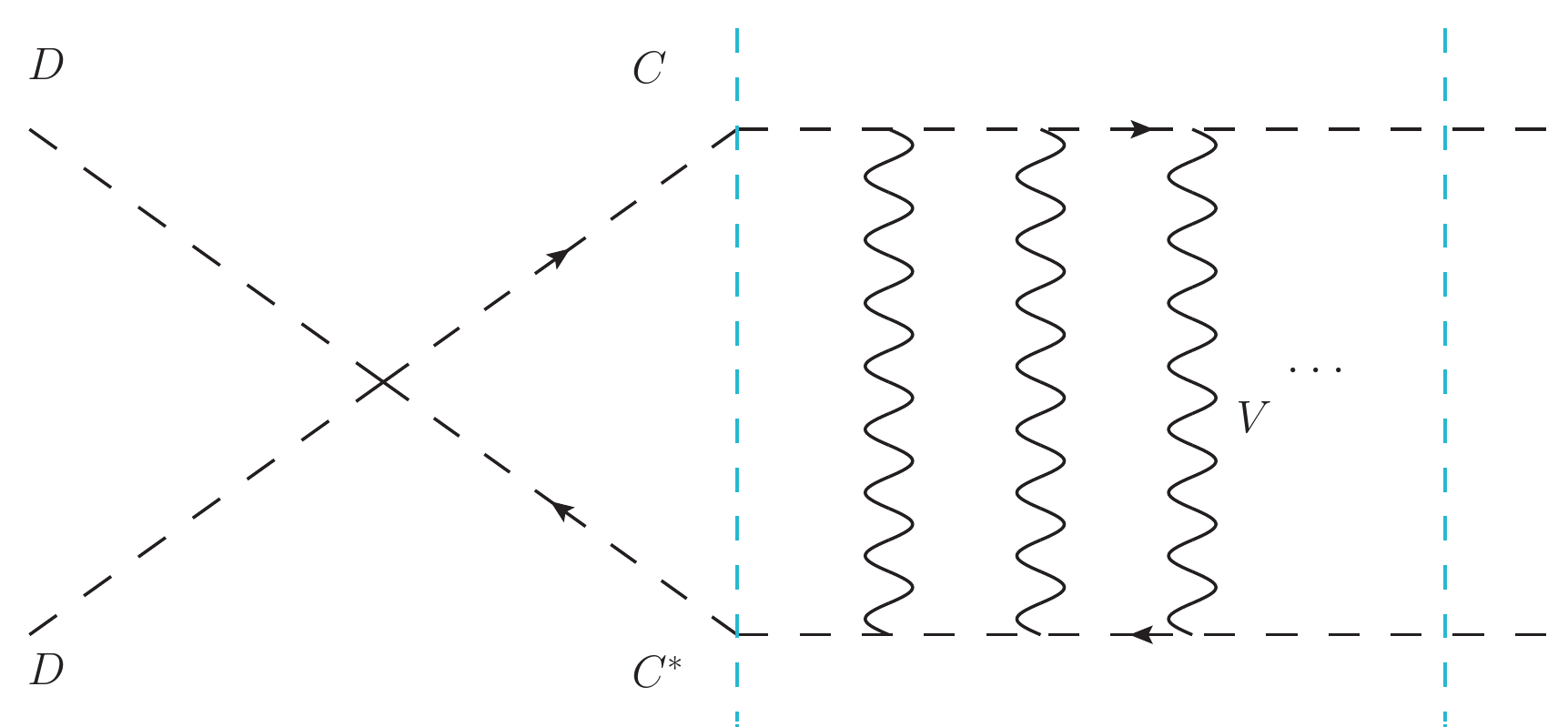}}
  \caption{\it Feynman diagrams for $DD\rightarrow CC^*$ annihilation without/with FSS effect.}
    \label{fig:SSF}
\end{figure}
The amplitude for this process is
\begin{equation}
    i{\mathcal{M}}_{DD\rightarrow CC^*} =-ig_D.
\end{equation} 
The cross section times relative velocity $v$ of the incoming DM particles in the (center-of-momentum) COM frame is
\begin{equation}
    \begin{aligned}
      & (\sigma _{ann} v) =  \frac{g_D^2}{8\pi s}v_2 \\
      & v_2=\sqrt{1-4m_C^2/s},
    \end{aligned}
\end{equation}
where $s=(p_1+p_2)^2$ is the Mandelstam variable; $v_2$ is the velocity of final state particle $C$ in COM. It is the part under the square root that has to be greater than $zero$ for ``forbidden'' case. We can directly calculate $s_{min}$ by setting it equal $zero$ for the ``forbidden'' case. Therefore, when $m_C>m_D$, $s_{min} = 4 m_C ^2 $; otherwise, $s_{min} = 4 m_D ^2 $.

Since the final state particle $C$ has a coupling with $V$, we shall consider a simple Coulomb-like potential between the final $CC^\ast$ pair in non-relativistic. The consequence is the FSS effect. This effect has been previously considered in \cite{Cui:2020ppc}. Because the matrix element is a constant here, there is only the $s$-wave Sommerfeld effect. Multiply the $s$-wave Sommerfeld factor, the cross section takes the form of
\begin{equation}
  \begin{aligned}
    & (\sigma v)_{FSS} = (\sigma _{ann} v) \ S_f \\
    & S_f=\frac{\pi\alpha_{V}/v_2}{1-e^{-\pi\alpha_{V}/v_2}}, \ \alpha_{V} = \frac{g_V ^2}{4 \pi}
  \end{aligned}
\end{equation}
where $S_f$ is the FSS factor for the $s$-wave annihilation.

For the FSS effect, since $S_f$ is applicable to a non-relativistic final state velocity, we choose to turn off the FSS effect for $v_2>0.6$. In the following calculation, we substitute $S_f$ by $[(S_f-1)H(0.6-v_2)+1]$, where $H(0.6-v_2)$ is the Heaviside step function following the treatment in \cite{Cui:2020ppc}.

After calculating the cross section and the $s_{min}$, one can follow the process in Section \ref{sec:Thermal averag} to work out the thermal- averaged $\langle\sigma v\rangle_{FSS}$.

\subsubsection{FBS formation without emission}

\begin{figure}[H]
  \centering
  \includegraphics[width=0.45\textwidth]{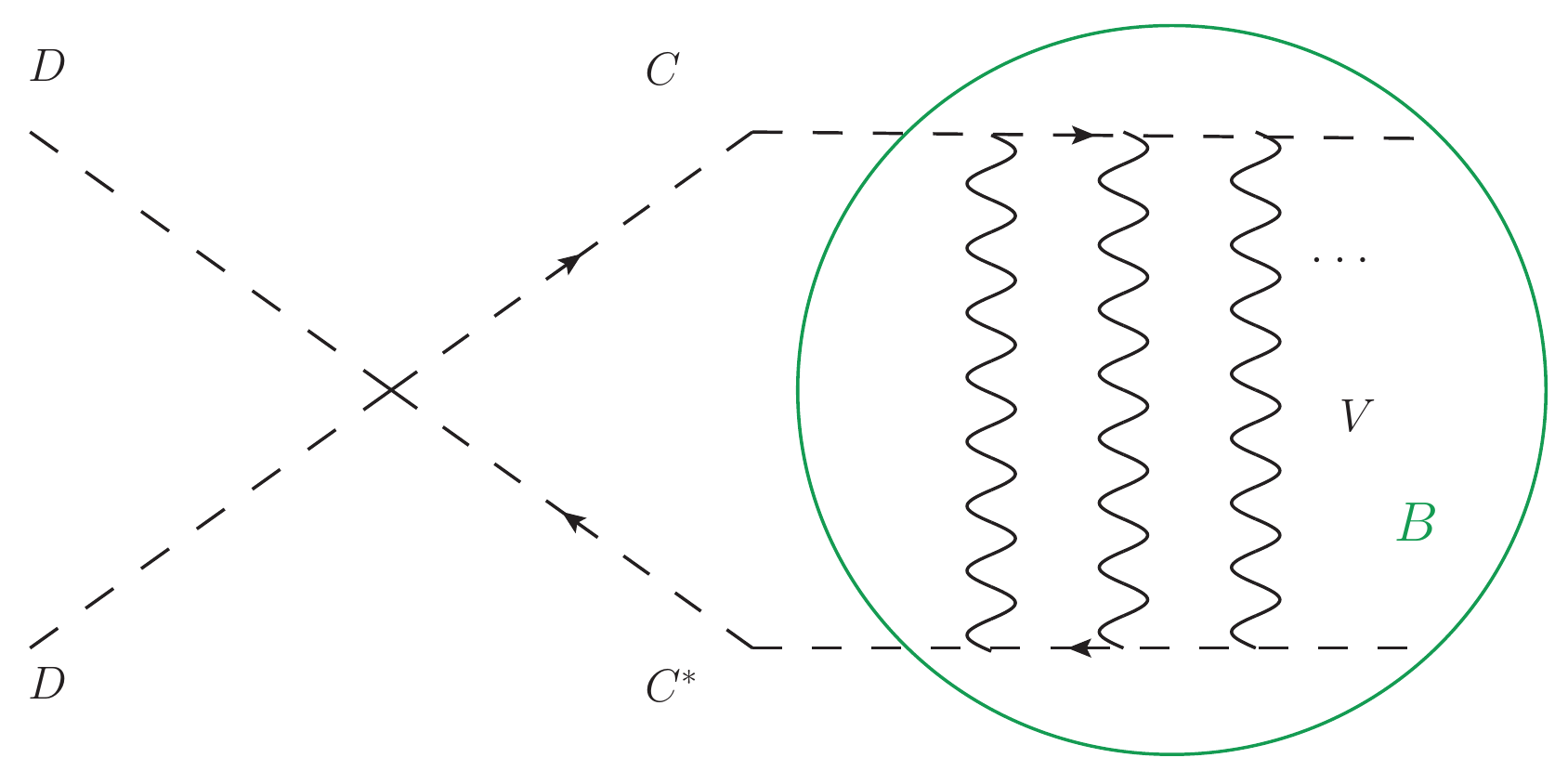}
  \caption{\label{fig:DDB} \it The Feynman diagram for the $DD^\ast \rightarrow B$ annihilation.}
\end{figure}

$C$ and $C^*$ can bind into positronium-like states, the $s$-wave (orbital angular momentum $L=0$) FBS can be allowed to form. Fig. \ref{fig:DDB} shows this process.

The scattering amplitude of this process considering the ground bound state formation is
\begin{equation}
  \mathcal{M} _{DD \rightarrow B(L=0)} = \sqrt{\frac{1}{m_C}}\int \frac{d^3\mathbf{k}}{(2\pi)^3} \widetilde{\psi}_{n00} ^*(\mathbf{k})\mathcal{M}_{DD\rightarrow CC^*}.
\end{equation}
 where $\widetilde{\psi}$ is the wave function in momentum space. Because the matrix element for $DD\rightarrow CC^*$ does not depend on $\mathbf{k}$, we can directly use the relation
\begin{equation}
   \int\frac{d^3\mathbf{k}}{(2\pi)^3}\widetilde{\psi}^*_{n00} (\mathbf{k})=\psi^*_{n00} (\mathbf{r}=0), 
\end{equation}
where
\begin{equation}
\label{psi}
   \psi^*_{n00}(\mathbf{r}=0)=\frac{1}{\sqrt{\pi}(na_0)^{3/2}}
\end{equation}
is the $s$-wave Hydrogen-like wave function for $CC^*$ bound state, where $a_0=1/(\mu\alpha_{V})$ is the Bohr radius, reduce mass $\mu=m_C/2$. 
We denote this process as $DD\rightarrow B(L=0)$, and work in the COM frame, then obtain
\begin{equation}
     |\mathcal{M}_{DD\rightarrow B(L=0)}|^2=\frac{g_D^2\alpha_{V}^3m_C^2}{8n^3\pi}
\end{equation}
The cross section times relative velocity is \cite{Peskin:1995ev}
\begin{equation}
  (\sigma v)_B = \frac{2\pi}{s}|\mathcal{M} _{DD\rightarrow B(L=0)}|^2  \ \delta \left(E_{cm}^2-m_B^2\right),
\end{equation}
where $m_B=2m_C-E_B=2m_c-\frac{\alpha_{V}^2m_C}{4n^2}$ is bound state mass, $E_B$ is the binding energy; $\delta$ function ensures energy-momentum conservation. Again, the thermal-averaged cross section $\langle\sigma v \rangle_{DD\rightarrow B}$ can be worked out according to Section \ref{sec:Thermal averag}.

In the numerical calculation, we only include the $n=1,L=0$ FBS. It is also possible to form FBS with the principal quantum number larger than $1$, corresponding to $\widetilde{\psi}^*_{n00} (\mathbf{k})$, and the result is amount to multiply a factor $\frac{1}{n^3}$ to $\psi^*_{100}(\mathbf{r}=0)$. The total contribution for these excited bound states will enlarge the result of FBS formation without emission by a factor of order $1$. We neglect those contribution in this work, and therefore our result is conservative. For other cross sections in Model I and Model II, we still only consider the relevant smallest principal quantum number.   

\subsubsection{FBS formation with emission}

\begin{figure}[H]
  \centering
  \subfigure{
  \includegraphics[width=0.45\textwidth]{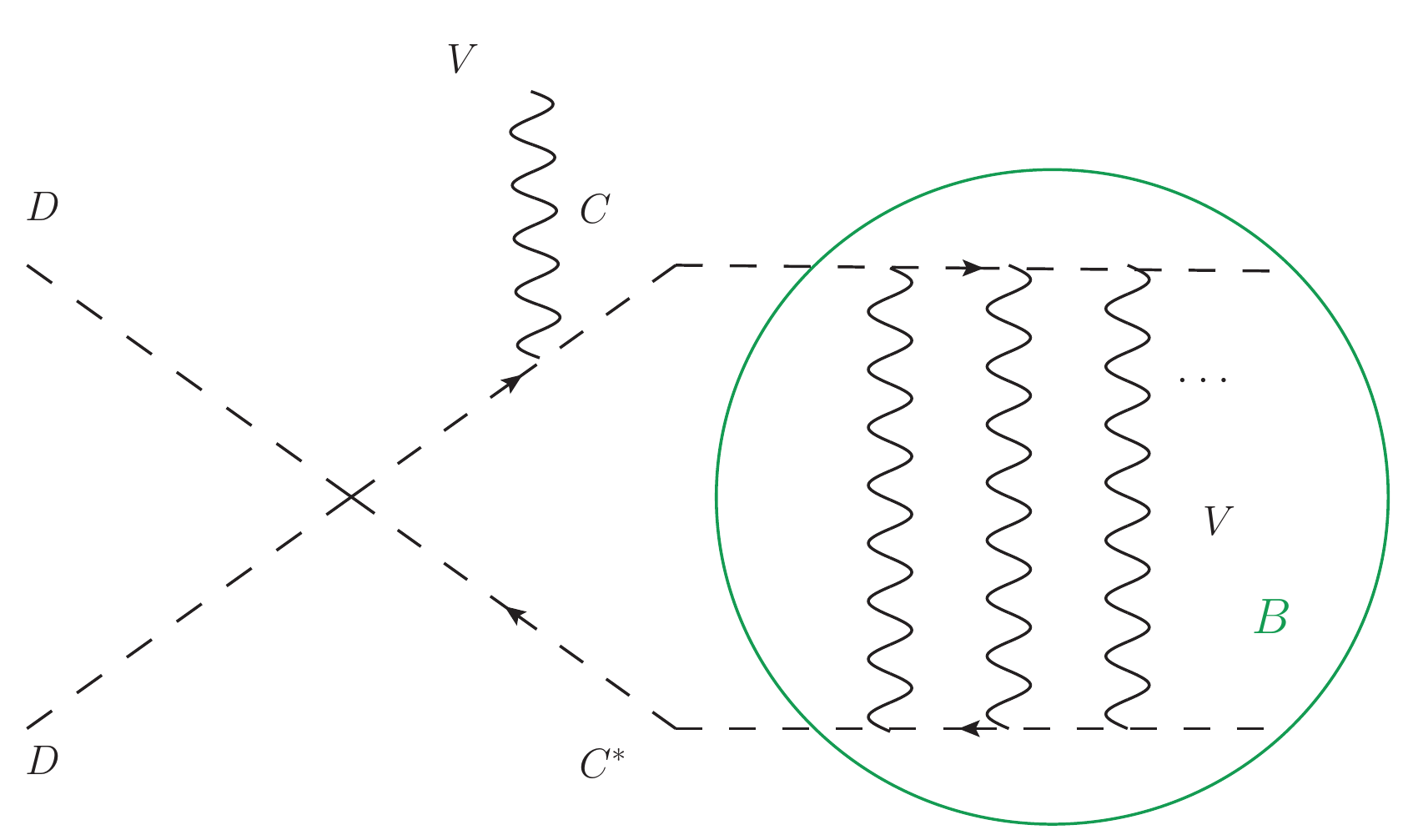}}
  \hspace{0in}
  \subfigure{
  \includegraphics[width=0.45\textwidth]{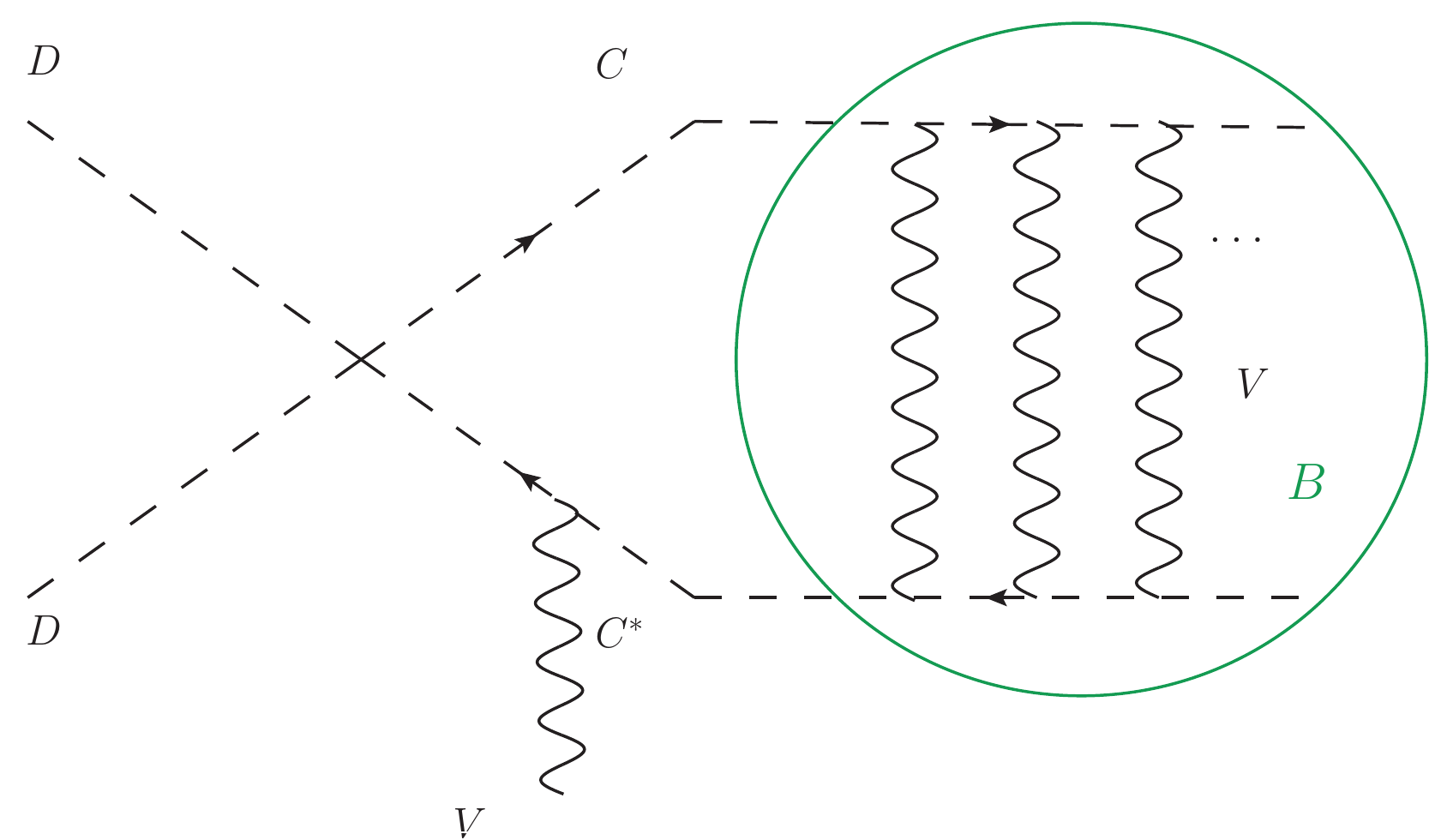}}
  \caption{\it The Feynman diagrams for the $(CC^*)$ bound state production process $DD\rightarrow BV$. }
    \label{fig:DDBV}
\end{figure}

Bound state can also form via a vector boson emission process as shown in Fig. \ref{fig:DDBV}. First of all, noting that vector boson carries spin one, the first term allowed is the $p$-wave (orbital angular momentum $L=1$) FBS. 

The scattering amplitude for the process $DD\rightarrow CC^*V$ can be written as
\begin{equation}
\label{eq:origin DD-CCV}
  \begin{aligned}
  &i\mathcal{M}_{DD\rightarrow CC^*V} = ig_Dg_V(\frac{2k_2\cdot \epsilon^*(q)+q\cdot\epsilon^*(q)}{2k_2\cdot q+m_V^2}-\frac{2k_1\cdot \epsilon^*(q)+q\cdot\epsilon^*(q)}{2k_1\cdot q+m_V^2}),
  \end{aligned}
\end{equation}
where $p_1$, $p_2$, $k_1$, $k_2$ and $q$ are respectively the momenta of $D$, $D$, $C$, $C^*$ and $V$. $\epsilon _\mu$ is the vector boson $V$ polarization vector satisfying
\begin{equation}
  \label{eq:8}
  \sum_{\lambda =1}^{3}\epsilon_\mu (q)\epsilon_\nu (q)=-g_{\mu\nu}+\frac{q_\mu q_\nu}{m_V^2}.
\end{equation} 

We introduce the total and relative momenta  $\mathbf{K}=\mathbf{k_1}+\mathbf{k_2}$ and $\mathbf{k}=(\mathbf{k_1}-\mathbf{k_2})/2$ for $C$ and $C^*$. In the bound state rest frame, $\mathbf{K}=0$. It allows us to rewrite Eq. (\ref{eq:origin DD-CCV}) under non-relativistic approximation and expand to the first order of $\mathbf{k}$, as
\begin{equation}
\mathcal{M}^j_{DD\rightarrow CC^*V}\simeq  -g_Dg_V\left(\frac{4k^j}{2m_C\omega +m_V^2}+\frac{4q^j(\mathbf{k}\cdot\mathbf{q})}{(2m_C\omega+m_V^2)^2}\right),
\end{equation} 
where $\omega$ is the vector boson energy, index $j$ stands for spatial 3-components.

The products of $DD$ can form bound states, we write the amplitude $\mathcal{M}_{DD\rightarrow BV}$ in terms of final $CC^*$ pair momentum $\mathbf{K}$, $\mathbf{k}$. The FBS effect can be calculated as follow: the matrix element multiplies the momentum space wave function ($L=1$), integrating out the relative momentum $\mathbf{k}$ \cite{Peskin:1995ev,Petraki:2015hla}.
\begin{equation}
    \begin{aligned}
          \mathcal{M}^j_{BV}=\sqrt{\frac{1}{m_C}}\int \frac{d^3\mathbf{k}}{(2\pi)^3}\widetilde{\psi}_{21m} ^*(\mathbf{k}) \ \mathcal{M}^j_{DD\rightarrow CC^*V},
    \end{aligned}
\end{equation}

A mathematical trick can be used in the integrals with respect to $\mathbf{k}$:
\begin{equation}
  \label{eq:11}
  \int \frac{d^3\mathbf{k}}{(2\pi)^3} k^j \widetilde{\psi} ^{*i}(\mathbf{k})=-i\nabla^j\psi ^i(\mathbf{x})|_{\mathbf{x}=0}.
\end{equation}
The value of $L=1$ wave function at $\mathbf{r}=0$ is $zero$. It's clear that bound state formation matrix elements are proportional to the value of the first-order derivative of $L=1$ wave function at $\mathbf{r}=0$. The details to deal with $p$-wave bound-state can be seen more in \cite{Asadi:2016ybp,Sen:2018wnx}. Hydrogen-like wave function of $CC^*$ bound-state has the form
\begin{equation}
  \label{eq:wave function}
  \psi  _{nlm}(\mathbf{r})=\left[\frac{1}{2n}\left(\frac{2}{na_0}\right)^3\frac{(n-l-1)!}{(n+l)!}\right]^{1/2}\left(\frac{2r}{na_0}\right)^l\mathrm{e}^{-r/na_0}L^{2l+1}_{n-l-1}\left(\frac{2r}{na_0}\right)Y_{lm}(\theta ,\phi ),
\end{equation}
and 
\begin{equation}
  L^m_n(x)=(n+m)!\sum_{k = 0}^{n}\frac{(-1)^k}{k!(n-k)!(k+m)!}x^k,   
\end{equation}
are associated Laguerre polynomials.

For the polarization summation of vector boson, we can follow the method in \cite{Petraki:2016cnz}
\begin{equation}
\label{eq:14}
    \begin{aligned}
    \sum_{m,\lambda}|\mathcal{M}_{DD\rightarrow B(L=1)V}|^2&=\left(-g_{\mu\nu}+\frac{q_\mu q_\nu}{m_V^2}\right)\mathcal{M}^\mu\mathcal{M}^{\nu*}\\
    &=\mathcal{M}^j\mathcal{M}^{j*}-\frac{|q^j\mathcal{M}^j|^2}{\mathbf{q}^2+m_V^2}.
    \end{aligned}    
\end{equation}
Notice that $q_\mu\mathcal{M}^\mu=0$. Therefore, the formula is the same for massive and massless vector boson.

The final result after polarization summation is
\begin{equation}
\label{eq:18}
  \sum_{m,\lambda}|\mathcal{M}_{DD\rightarrow B(L=1)V}|^2=C\left[\left(3-B\right)+2A\left(1-B\right)+A^2\left(1-B\right)\right],
\end{equation}
where
\begin{equation}
\begin{aligned}
A=&\frac{|\mathbf{q}|^2}{2m_C\omega+m_V^2}=\frac{\omega^2-m_V^2}{2m_C\omega+m_V^2}\\
B=&\frac{|\mathbf{q}|^2}{|\mathbf{q}|^2+m_V^2}\\
C=&\frac{g_D^2}{6}\frac{n^2-1}{n^5}\alpha_{V}^6 \frac{4m_C^4}{(2m_C\omega+m_V^2)^2}.
\end{aligned}
\end{equation}
In the bound state rest frame, the emission vector boson energy $\omega$ and 3-momentum modulus square already fixed by mandelstam variable $s$ and the masses of particles
\begin{equation}
    \begin{aligned}
         & \omega = \frac{s-m_B^2-m_V^2}{2m_B}\\
          & |\mathbf{q}|^2 = \omega^2 - m_V^2.
    \end{aligned}
\end{equation}

It is obvious that the boson momentum $|\mathbf{q}|$ must be larger than $zero$, it decides the $s_{min}$. In the low temperature limit, $m_V \sim g T = 0$, only the first term in the square brackets is left in Eq. \eqref{eq:18}
\begin{equation}
    \sum_{m,\lambda}|\mathcal{M}_{DD\rightarrow B(L=1)V}|^2= \frac{g_D^2}{3}\frac{n^2-1}{n^5}\alpha_{V}^6 \left(\frac{m_C}{\omega}\right)^2.
\end{equation}
There is a pole when the energy of vector boson $\omega=0$, which is usual infrared divergence for soft bremsstrahlung. We can address this problem by introducing Debye mass as Eq.\eqref{eq:Debye mass} shown. 
In order to take care of these infrared divergence appropriately, we need loop diagrams to offset it. 

Because the $|\mathcal{M}|^2$ is Lorentz invariant, for simplicity, we change the reference frame to COM frame for the phase space integration according to the general formula \cite{Peskin:1995ev}, the cross section times relative velocity $v$ of the incoming DM particles for the process $DD\rightarrow B(L=1)V$ gives
\begin{equation}
\label{eq:qcm}
\begin{aligned}
    & (\sigma v)_{BV}=\frac{\sum_{m,\lambda}|\mathcal{M}_{DD\rightarrow B(L=1)V}|^2}{4 \pi s } \frac{|\mathbf{q}|_{cm}}{\sqrt{s}} \\
    & |\mathbf{q}|_{cm} = \sqrt{\left(\frac{s+m_B^2-m_V^2}{2\sqrt{s}}\right)^2-m_B^2}.
\end{aligned}
\end{equation}

\subsection{Numerical results of Model I}

\label{sec:numerical result Model I}
In order to show the BSF formation effects in the non-relativistic region,  we plot the thermal-averaged cross sections at three parameters $\alpha_{V} = 0.02, 0.1, 0.5$ which represent electroweak-like, strong-like and ``super'' strong-like coupling, respectively. To explore the FBS effect in the non-relativistic region of final products, we fix the mass of annihilated DM and normalize other particles mass by it. We use $z$ as the ratio of product particle mass and DM mass, $z\equiv m_C/m_D$, then we plot the thermal-averaged cross sections evolution as a function of $x$ at three parameters $z = 0.9, 1, 1.1$. 

\subsubsection{Thermal averaged cross section}
\label{Thermal averaged cross section I}
The cross sections and kinematic threshold $s_{min}$ for three processes are summarised in Table. \ref{table:cross sections I}. In the numerical calculation, we only consider $n=1$ for the $DD\rightarrow B$ and $n=2$ for the $DD\rightarrow BV$.
\begin{table}[H]
  \centering
  \caption{Cross sections and kinematically forbidden limits for Model I.}
  \begin{tabular}{|l|c|c|c|}
  \hline
    $channel$ & $(\sigma v)$ &  $s_{min}$ \\
  \hline 
  $DD\rightarrow CC^*$ & $\frac{g_D^2}{8\pi s}v_2\frac{\pi\alpha_{V}/v_2}{1-e^{-\pi\alpha_{V}/v_2}}$ & $Max[4 m_D^2,4m_C^2]$ \\
  $DD\rightarrow B$ & $\frac{g_D^2\alpha_{V}^3m_C^2}{4s}\delta(E_{cm}^2-m_B^2)$ & $4 m_D^2$\\ 
  $DD\rightarrow BV$ & $\frac{|\mathbf{q}|_{cm}}{\sqrt{s}} \ C\left[\left(3-B\right)+2A\left(1-B\right)+A^2\left(1-B\right)\right]/(4\pi s)$ & $Max[4 m_D^2, \ (m_B+m_V)^2]$ \\
  \hline
  \end{tabular}
   \label{table:cross sections I}
\end{table}

\begin{figure}[H]
  \centering
   \subfigure{
  \includegraphics[width=0.3\textwidth]{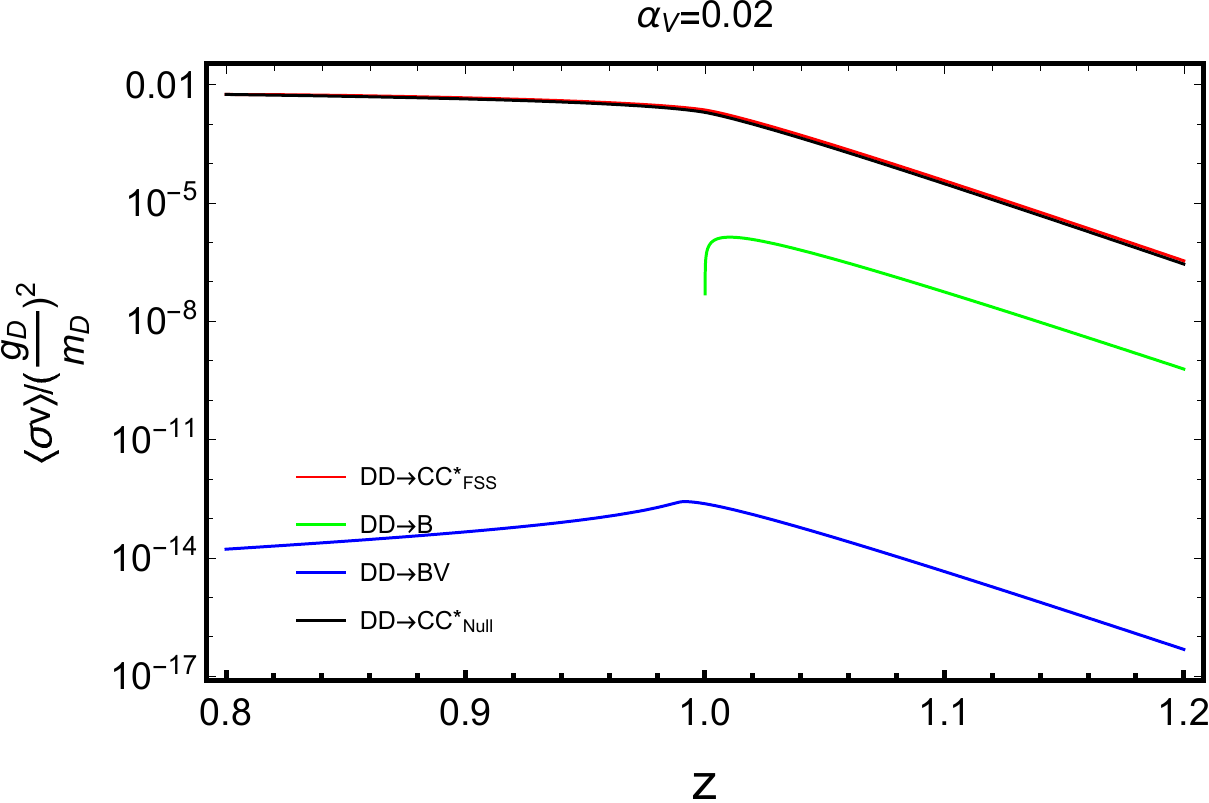}}
  \hspace{0in}
  \subfigure{
  \includegraphics[width=0.3\textwidth]{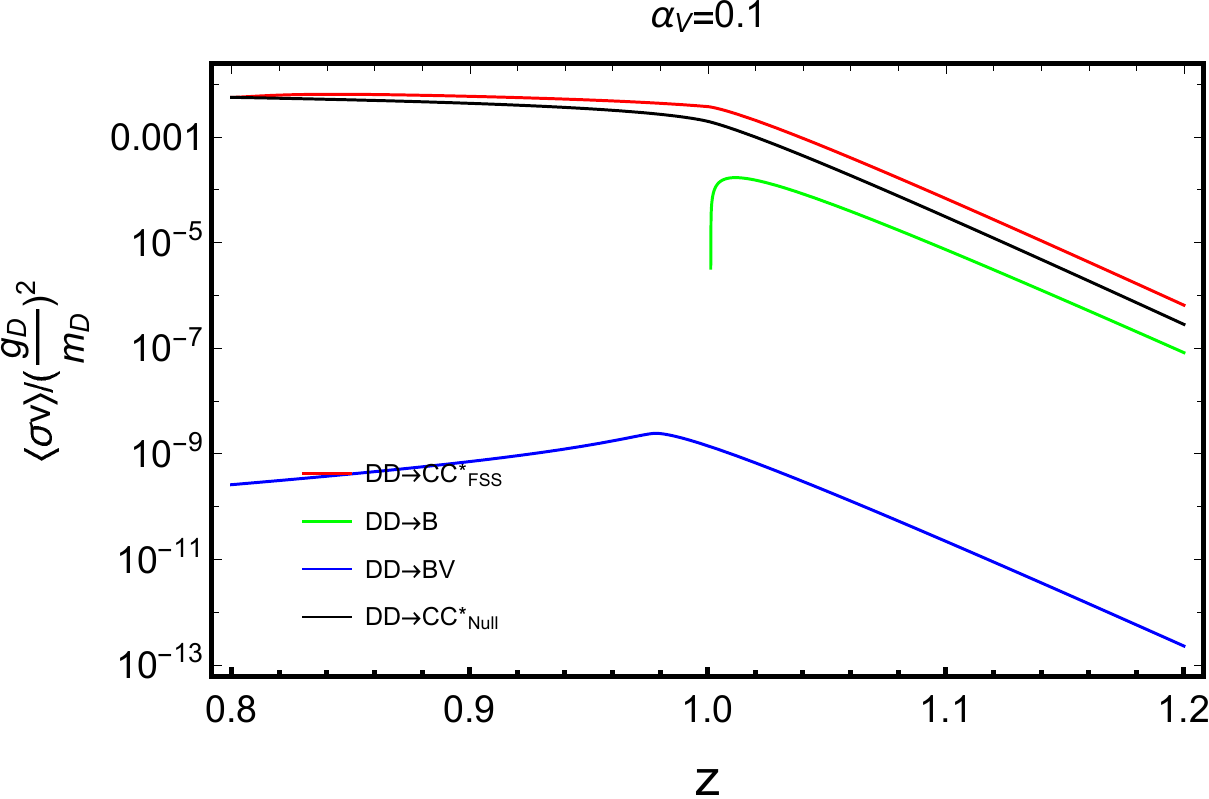}}
  \hspace{0in}
  \subfigure{
  \includegraphics[width=0.3\textwidth]{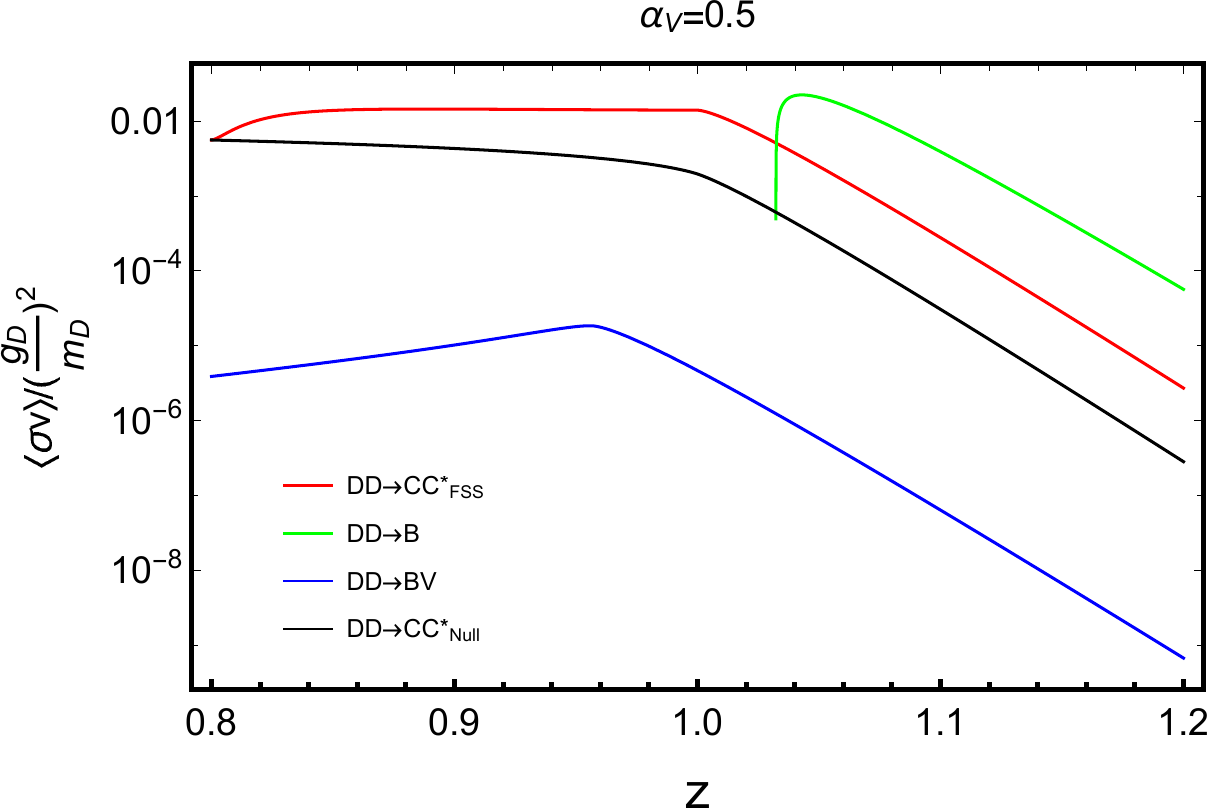}}
  \caption{\label{fig:thermal-averaged cross sections Model I}\it The thermal-averaged cross sections over a common factor $ g_D ^2 / m_D ^2$ at three parameters $\alpha_{V} = 0.02, 0.1, 0.5 $, at a typical freeze-out value $x=m_D/T=25$. The red, green, blue and black lines stand for $\langle\sigma v\rangle_{FSS}$, the thermal-averaged FSS-corrected $s$-wave cross section; $\langle\sigma v \rangle_{B}$, the thermal-averaged $s$-wave FBS (without boson emission) formation cross section; $\langle\sigma v \rangle_{BV}$, the thermal-averaged $p$-wave FBS (with boson emission) formation cross section; and $\langle\sigma v \rangle_{w/o \ both}$, the thermal-averaged cross section without any FSS and FBS, respectively. $z$ is the mass ratio $m_C/m_D$; the $y$-axis is the thermal-averaged cross sections divided by a common factor. }
\end{figure}

We can directly calculate the thermal-averaged cross sections following the Eq. \eqref{average we use} in Section \ref{sec:Thermal averag}. Fig. \ref{fig:thermal-averaged cross sections Model I} shows the thermal-averaged cross sections, we choose three values of $\alpha_{V}$, $0.02$, $0.1$ and $0.5$ for illustration, indicating electroweak-like, strong-like and ``super'' strong-like couplings, respectively. The red, green, blue and black lines stand for $\langle\sigma v\rangle_{FSS}$, $\langle\sigma v \rangle_{B}$, $\langle\sigma v \rangle_{BV}$ and $\langle\sigma v \rangle_{w/o \ both}$ over a common fator $ g_D ^2 / m_D ^2$, respectively, as functions of $z$ at a typical freeze-out value $x=25$. The subscript ``w/o  both'' indicates that both FSS and FBS effect are not included.

For a given $\alpha_{V}$, the $\langle\sigma v\rangle_{B}$ gives much more contribution than $\langle\sigma v\rangle_{BV}$, because in Model I, $\langle\sigma v\rangle_{BV}$ corresponding to $p$-wave FBS formation, are suppressed by $\alpha_{V}^3$ (two from the square of wave function and one from interaction vertex). When $z>1$, single $s$-wave bound state can be formed\footnote{The $DD\to B$ process starts from $2m_D=m_B=2m_C-E_{B}$, where $E_{B}$ is the binding energy. So the green lines start from $z>1$, where $z=m_C/m_D$.} without vector boson emission, so $\langle\sigma v\rangle_{B}$ becomes comparable with $\langle\sigma v\rangle_{FSS}$ for large $\alpha_{V}$. In Fig. \ref{fig:thermal-averaged cross sections Model I}, as $\alpha_{V}$ increases from left to right, the binding energy of FBS increases meaning it is easier to form bound state  and the bound state gets tighter. The FSS effect also increases because the same light mediator mediates long range force.  

The different partial wave FSS/FBS effect is sensitive to the order of $\alpha_{V}$. From above figure we can read out the difference of partial wave contributions, and this partial wave FSS/FBS effect relative contributions also change significantly as $\alpha_{V}$ value changes.

\begin{figure}[H]
  \centering
   \subfigure{
  \includegraphics[width=0.3\textwidth]{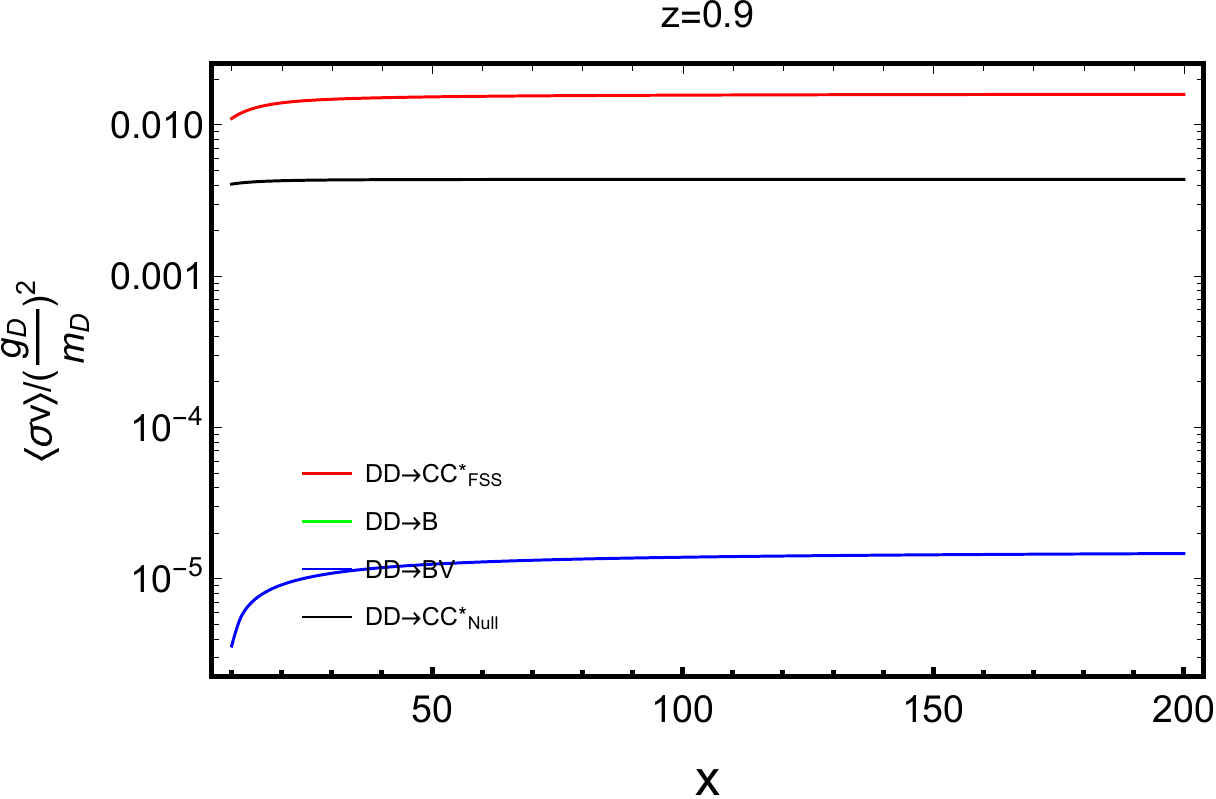}}
  \hspace{0in}
  \subfigure{
  \includegraphics[width=0.3\textwidth]{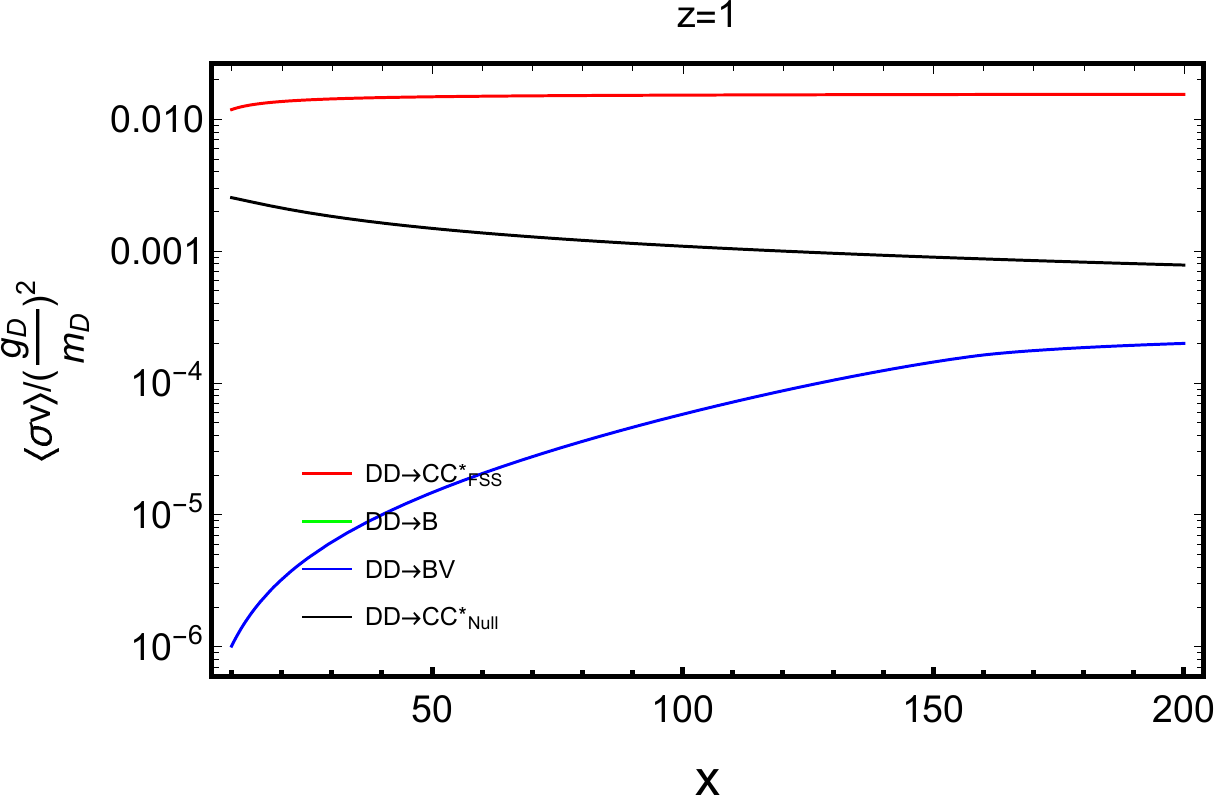}}
  \hspace{0in}
  \subfigure{
  \includegraphics[width=0.3\textwidth]{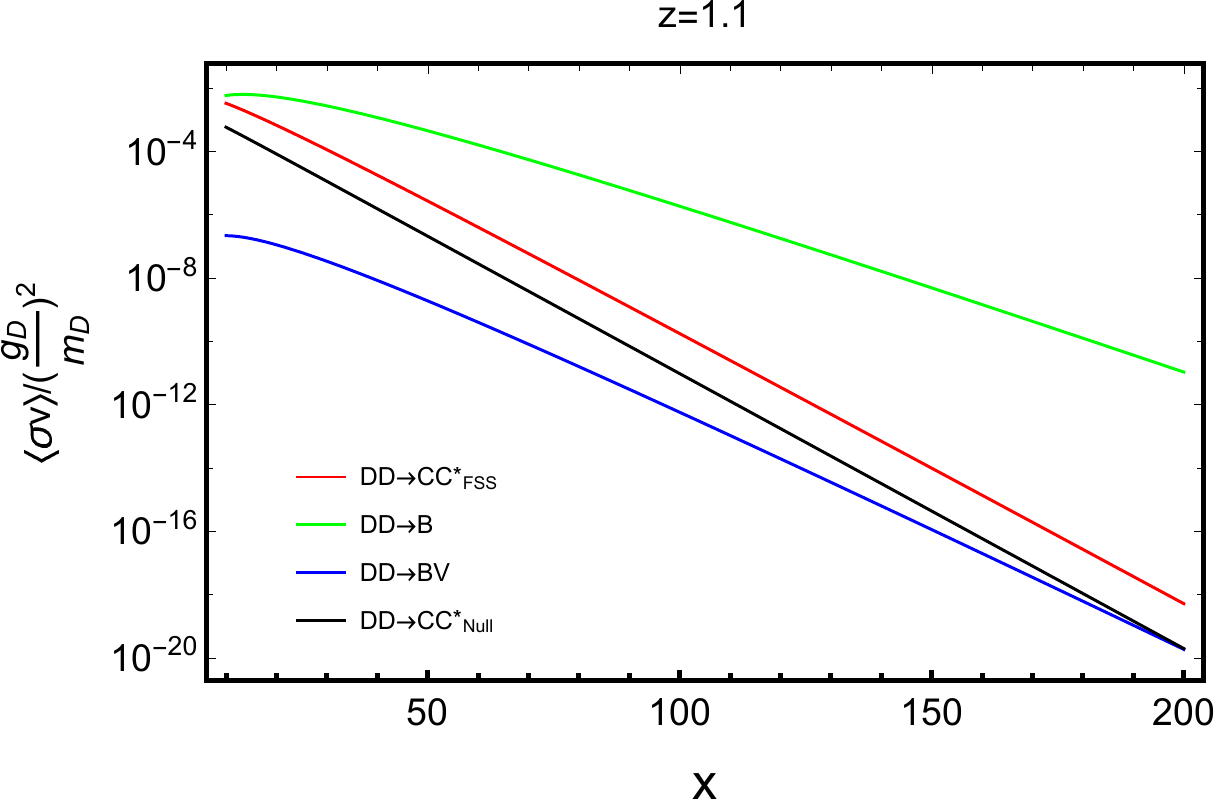}}
  \caption{\label{fig:thermal-averaged cross sections Model I,f(x)}\it The thermal-averaged cross sections over a common factor $ g_D ^2 / m_D ^2$ at three parameters $z = 0.9, 1, 1.1 $ and $\alpha_{V}=0.5$. The red, green, blue and black lines stand for $\langle\sigma v\rangle_{FSS}$, the thermal-averaged FSS-corrected $s$-wave cross section; $\langle\sigma v \rangle_{B}$, the thermal-averaged $s$-wave FBS (without boson emission) formation cross section; $\langle\sigma v \rangle_{BV}$, the thermal-averaged $p$-wave FBS (with boson emission) formation cross section; and $\langle\sigma v \rangle_{w/o \ both}$, the thermal-averaged cross section without any correction, respectively. The $x$ is the ratio $m_D/T$; the $y$-axis is the thermal-averaged cross sections divided by a common factor. }
\end{figure}

Fig. \ref{fig:thermal-averaged cross sections Model I,f(x)} shows the thermal-averaged cross sections evolution as temperature cools down. The missing of $\langle\sigma v \rangle_{B}$ in the left and the middle panel is because the initial energy in $z=0.9$ and $z=1$, are always larger than the single bound state energy, the rest mass of forming bound state must be released by emitting a vector boson; so the FBS effect is only contributed by $DD \to BV$ process.

Furthermore, in Fig. \ref{fig:thermal-averaged cross sections Model I,f(x)}, there is some difference between left/middle panel from the right panel. The right panel represents the forbidden case, as the $x$ increases, the thermal-averaged cross sections become smaller. Because as the temperature decreases, the initial particles become less energetic, the proportion of particles reaching the reaction threshold reduces. While, in the left and middle panel, cross sections do not change much with temperature. Because the Modle I is typical four point interaction and the scattering matrix element does not relate with the initial particles momentum.

The bound state can also be virtual as propagator in the $DD \to B^* \to V V$  process. The corresponding corss section can be calculated by Breit–Wigner formula \cite{Ibe:2008ye,Guo:2009aj}. However, we do not need to consider the virtual bound state process here. Because for $z>1$, the bound state, generated by the process of $DD \to B$, will decay and cause double counting; for $z<1$, in the Breit–Wigner formula, the virtual bound process is suppressed by the factor $\frac{m_B^{2} \Gamma^{2}}{\left(s-m_B^{2}\right)^{2}+m_B^{2} \Gamma^{2}}$, where $\Gamma \sim \alpha_{V}^5 m_B$, $m_B \approx 2 m_C$ (assuming the dominate decay channel is $B \to VV$). At the typical DM freeze-out temperature $x=25$, $(s-m_B^2) \sim \frac{12}{25} m_C^2$ at $z=1$, this factor is far away from its resonance pole.

\subsubsection{Relic abundance}

\begin{figure}[H]
  \centering
   \subfigure{
  \includegraphics[width=0.3\textwidth]{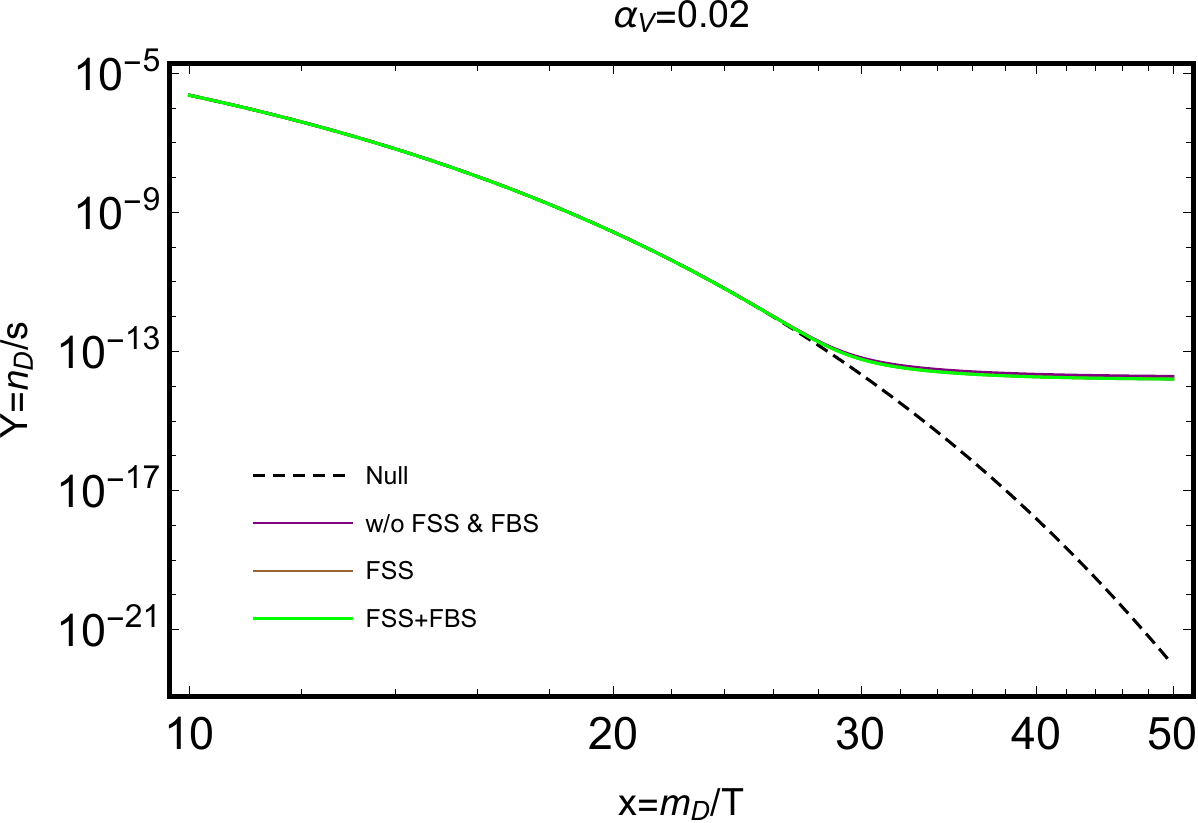}}
  \hspace{0in}
  \subfigure{
  \includegraphics[width=0.3\textwidth]{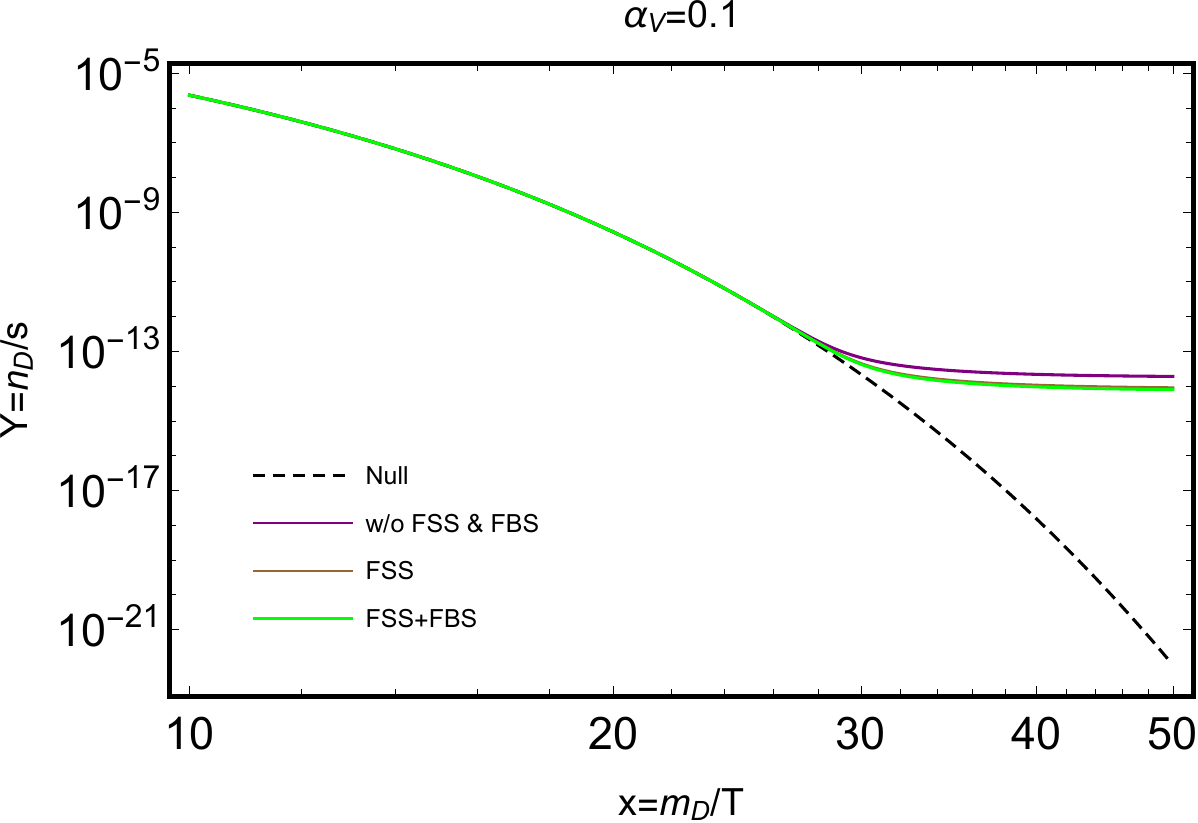}}
  \hspace{0in}
  \subfigure{
  \includegraphics[width=0.3\textwidth]{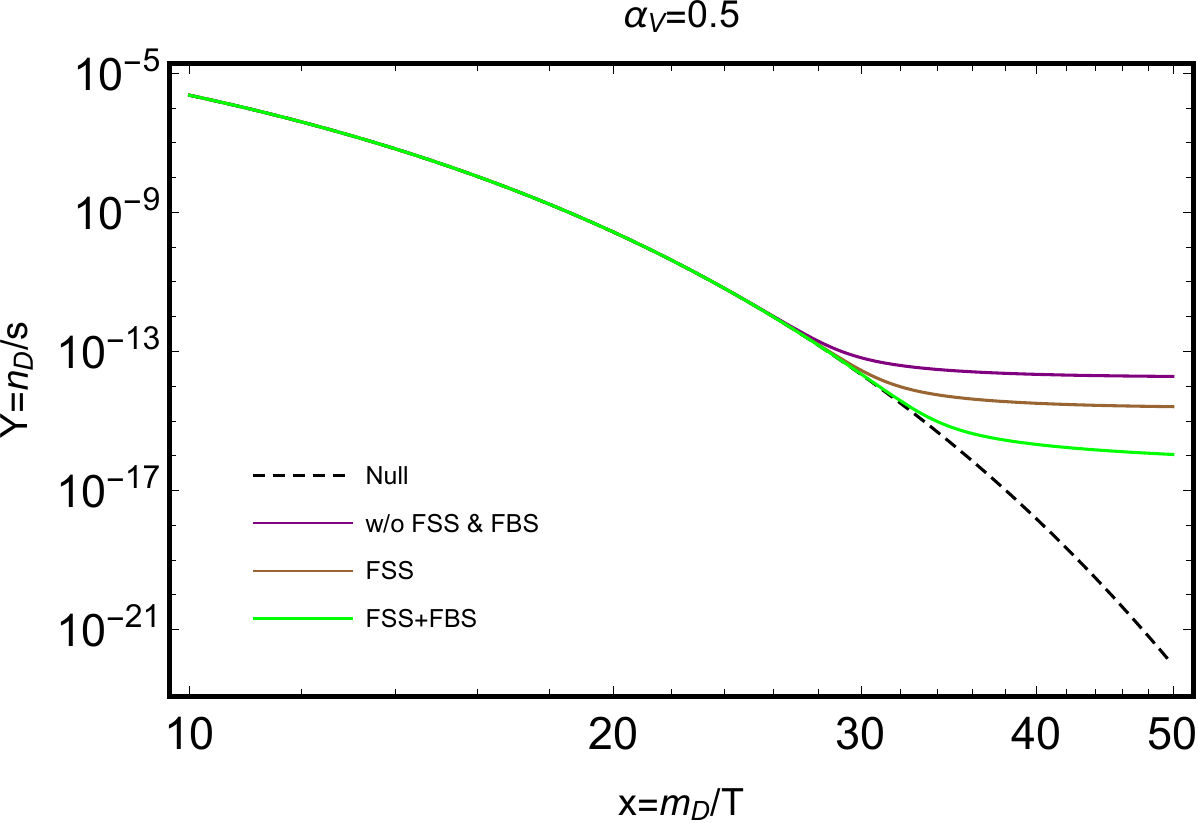}}
  \caption{\label{fig:model_I_abundance1}\it The evolution of the DM $yield$ as a function of $x=m_D/T$ for the representative case $m_D=1 TeV$, $z=1.1$, $g_D=1$ and $\alpha_{V}=0.02, 0.1, 0.5$. The purple line neglects both FSS and FBS effects. The brown line neglects FBS effect. The green line incorporates the effect of FBS and FSS. The black dashed line exhibits the naive thermal equilibrium abundance. }
\end{figure}

We already obtained the thermal-averaged cross sections numerically in \ref{Thermal averaged cross section I}, it is not hard to solve the Boltzmann equation as Section \ref{sec:Boltzmann equation} outlined. Similarly, we choose three parameters $\alpha_{V} = 0.02, 0.1, 0.5 $, and show the $yield$ of DM as a function of $x$, considering different effects. In order to show the FBS formation effect we choose other parameters as, $z=1.1$ (the forbidden case), and $m_D=1TeV$, $g_D=1$. As Fig. \ref{fig:model_I_abundance1} shows, the purple line neglects both FSS and FBS effects, the brown line neglects FBS effect, the green line incorporates the effects of FBS and FSS. In both Model I and Model II numerical calculation, we choose other parameters in the Boltzmann equation as $\frac{dg_{*s}}{dT}\simeq 0$, $g_*=g_{*s}=108.75$, We take $108.75$ to account for the SM plus the two dark-photon degrees of freedom.

Fig. \ref{fig:model_I_abundance1} shows the DM $yield$ considering different effects, with the coupling constant $\alpha_{V} = 0.02, 0.1, 0.5$, respectively. In the left panel, nothing surprise, for the electroweak-like scale interaction, $\alpha_{V}=0.02$, both FSS and FBS effects hardly change the DM $yield$. As the $\alpha_{V}$ increases to $0.1$, the FSS effect starts to show some influence, but the FBS effect is still feeble compared with FSS effect. In the right panel, $\alpha_{V}$ increases to $0.5$, the FBS shows a significant enhancement on the final $yield$ of DM, it further reduces the relic abundance by 93 $\%$ on top of the FSS effect. We note that for the forbidden case ($z=1.1$), all $\langle\sigma v \rangle$'s quickly decrease with the decrease of the temperature, as shown in the right panel of Fig. \ref{fig:thermal-averaged cross sections Model I,f(x)}. Thus, the DM $yield$ quickly reaches its asymptotic value after freeze-out. We have checked that the $yield$ is nearly the same at $x=50$ and $x=300$.

Fig.\ref{fig:thermal-averaged cross sections Model I} and \ref{fig:model_I_abundance1} show that the FBS formation effect and FSS effect are important in DM relic abundance calculation when DM annihilation products are non-relativistic and have a large coupling with a light vector boson. In particular, when $\alpha_{V}$ is very large, the cross section of FBS formation without emission dominates for the forbidden region.

\section{Model II}
\label{sec:Model II}

Model I is a typical four-point interaction between DM and the annihilation products. Due to angular momentum conservation, in Model I, we only consider $s$-wave FSS effect, $s$-wave FBS formation without vector boson emission, and $p$-wave FBS formation with a vector boson emission. Next we employ another model which at the leading order gives $p$-wave FSS effect, $p$-wave FBS formation without vector boson emission, and $s$-wave FBS formation with a vector boson emission.

In this model, DM consists of a complex scalar $S$, which has scalar QED-like coupling with a heavy neutral vector boson which we denote as $Z^\prime$ (but note that it is not the $Z$ boson in Standard Model). Another complex scalar $C$ couples with $Z^\prime$ and another light vector boson $V$. The dark sector we explore in this model is summarised by the Lagrangian: 
\begin{equation}
\label{eq:2}
    \mathcal{L}_{II}\supset|D_\mu C|^2+|D_\mu S|^2,
\end{equation}
where the covariant derivatives are $D^\mu C=\partial^\mu C+ig_3V^\mu C+ig_5Z ^{\prime \mu} C$, $D^\mu S=\partial^\mu S+ig_6Z^{\prime \mu} S$. 

At $zero$ temperature, the $V$ mass is zero. In the thermal bath, same as in Model I, it has Debye mass $m_V \sim g_VT $.

\subsection{Cross sections for Model II}
\subsubsection{Direct annihilation with FSS effect}

\begin{figure}[H]
  \centering
  \subfigure{
  \includegraphics[width=0.35\textwidth]{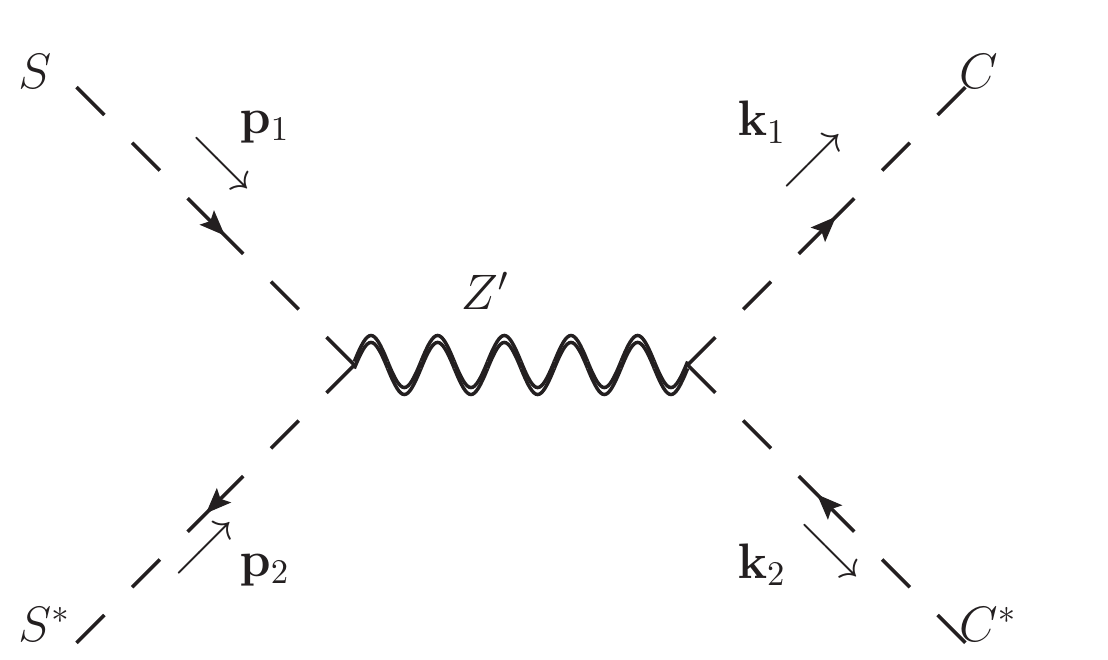}}
  \hspace{0in}
  \subfigure{
  \includegraphics[width=0.5\textwidth]{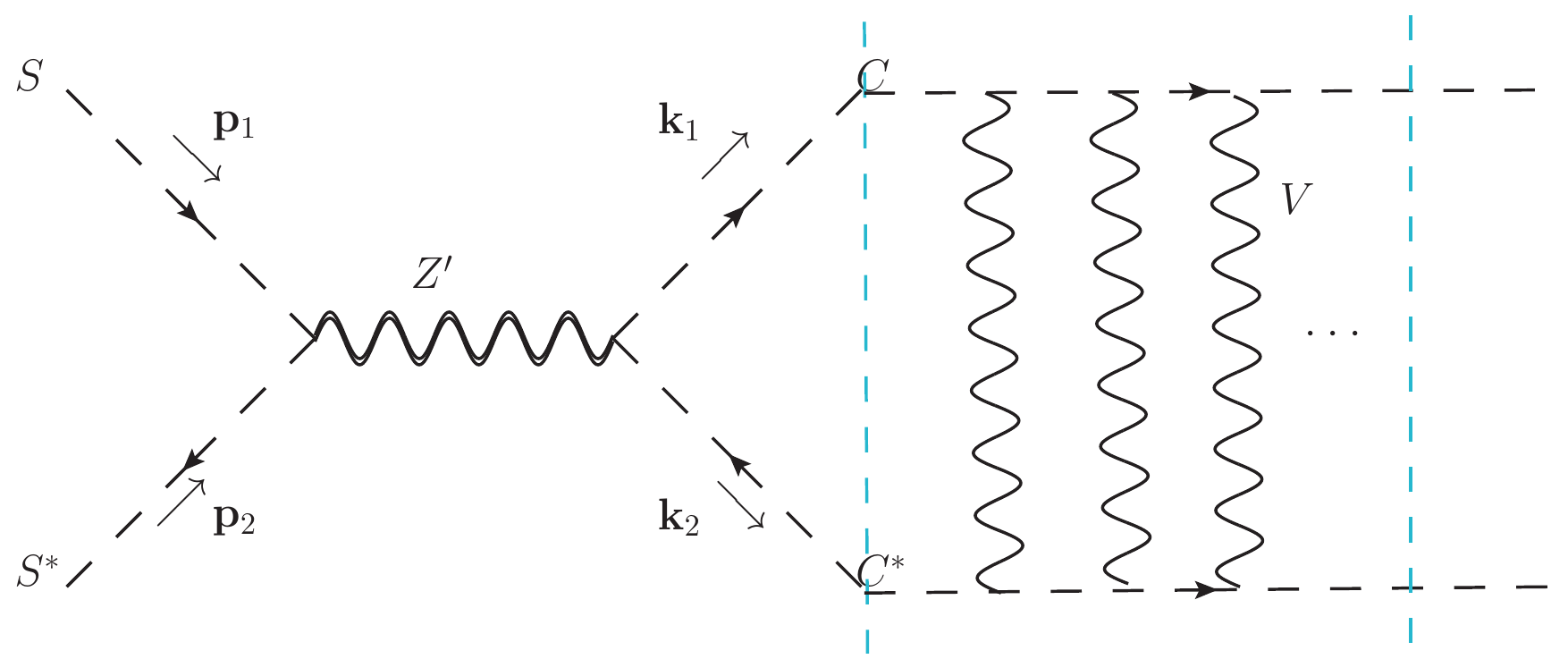}}
  \caption{\label{fig:SSCC} \it The Feynman diagrams for the $SS^\ast \rightarrow CC^\ast$ without/with FSS effect.}
\end{figure}

In this model, Fig. \ref{fig:SSCC} shows the direct annihilation of DM without/with FSS effect.

The scattering amplitude for the process $SS^*\rightarrow CC^*$ in the COM frame is
\begin{equation} \label{eq:SS-CC}
   i\mathcal{M}_{SS^*\rightarrow CC^*} = -4i g_5 g_6 \frac{|\mathbf{p}_1||\mathbf{k}_1|\cos\theta }{s-m_{Z^\prime}^2}.
\end{equation}
where $\mathbf{p}_1,\mathbf{k}_1$ are the 3-momentum, $s=(p_1+p_2)^2$. Because the matrix element is proportional to the final 3-momentum $\mathbf{k}_1$, there is only the $p$-wave Sommerfeld effect.

The cross section times relative velocity $v$ of the incoming DM particles in the COM frame for this process is
\begin{equation}
    \begin{aligned}
      & (\sigma _{ann}v) =\frac{g_5 ^2 g_6 ^2}{24 \pi s} \frac{(s-4m_S^2)(s-4m_C ^2)}{(s-m_{Z^\prime}^2)^2} v_2 \\
      & v_2=\sqrt{1-4m_C^2/s}.
    \end{aligned}
\end{equation}
Again, the value under the square root must be larger than $zero$ for the ``forbidden'' case.

Same as the Model I, the FSS effect can occur since the final state particles exchange vector boson $V$. We still use the Coulomb-like potential to calculate the FSS and FBS effect. Considering the FSS effect for $p$-wave, the corrected cross section is 
\begin{equation}
       (\sigma v)_{FSS}= (\sigma _{ann}v) \ S_f ,
\end{equation}
where
\begin{equation}
      S_f=(1+(\frac{\alpha_{V}}{2v_2})^2)\frac{\pi\alpha_{V}/v_2}{1-e^{-\pi\alpha_{V}/v_2}}
\end{equation}
is the FSS factor $S_f$ is for the $p$-wave \cite{Iengo:2009ni,Cassel:2009wt}, $\alpha_{V}=\frac{g_3^2}{4\pi}$.

\subsubsection{FBS formation without emission}

We get the scattering amplitude in Eq.\eqref{eq:SS-CC} of $SS^*\rightarrow CC^\ast$. As the last section outlined, the matrix element shows that the annihilation products should be in $p$-wave. And we have already discussed the $p$-wave FBS formation in Model I, and we use the same techniques to calculate the cross section for $SS^*\rightarrow B$ in Model II.
\begin{figure}[H]
  \centering
  \includegraphics[width=0.45\textwidth]{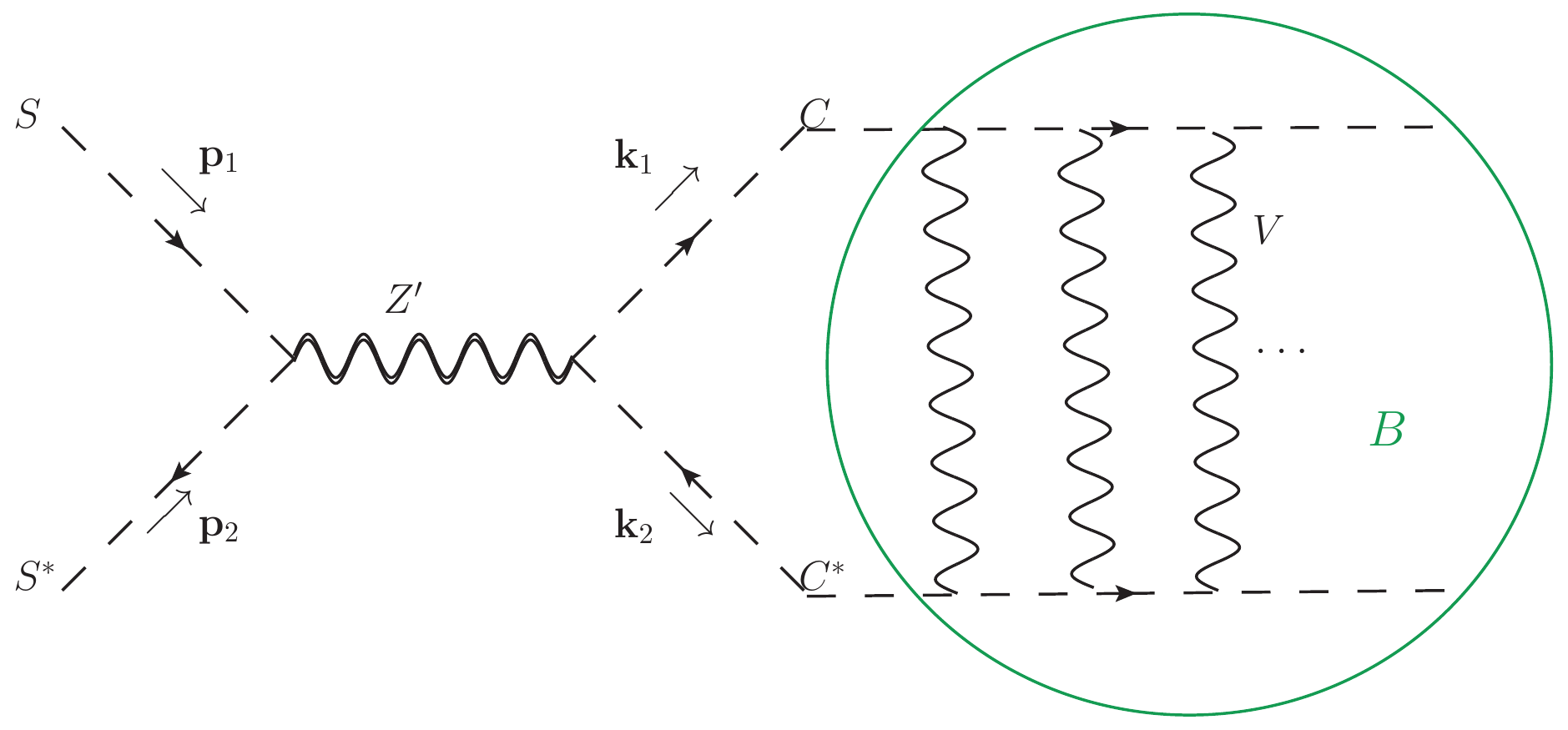}
  \caption{\label{fig:SSB} \it The Feynman diagrams for the $SS^\ast \rightarrow B$ annihilation.}
\end{figure}

Using the $L=1$ wave function in Eq. \eqref{eq:wave function} and working in the COM frame, we obtain
\begin{equation}
  |\mathcal{M}_{SS^*\rightarrow B}|^2=\frac{\alpha_{V} ^5 g_5^2g_6^2 }{24\pi}\frac{m_C^4 (s-4m_S^2)}{(s-m_{Z^\prime}^2)^2} (\frac{1}{n^3}-\frac{1}{n^5}).
\end{equation}
Then we get the cross section times relative velocity according to the \cite{Peskin:1995ev}
\begin{equation}
  (\sigma v)_B = \frac{2\pi}{s} |\mathcal{M}_{SS^*\rightarrow B}|^2 \ \delta \left(E_{cm}^2 - m_B^2\right),
\end{equation}
where $m_B$ is bound state mass, $\delta$ function ensures energy-momentum conservation.

\subsubsection{FBS formation with emission}

\begin{figure}[H]
  \centering
  \subfigure{
  \includegraphics[width=0.4\textwidth]{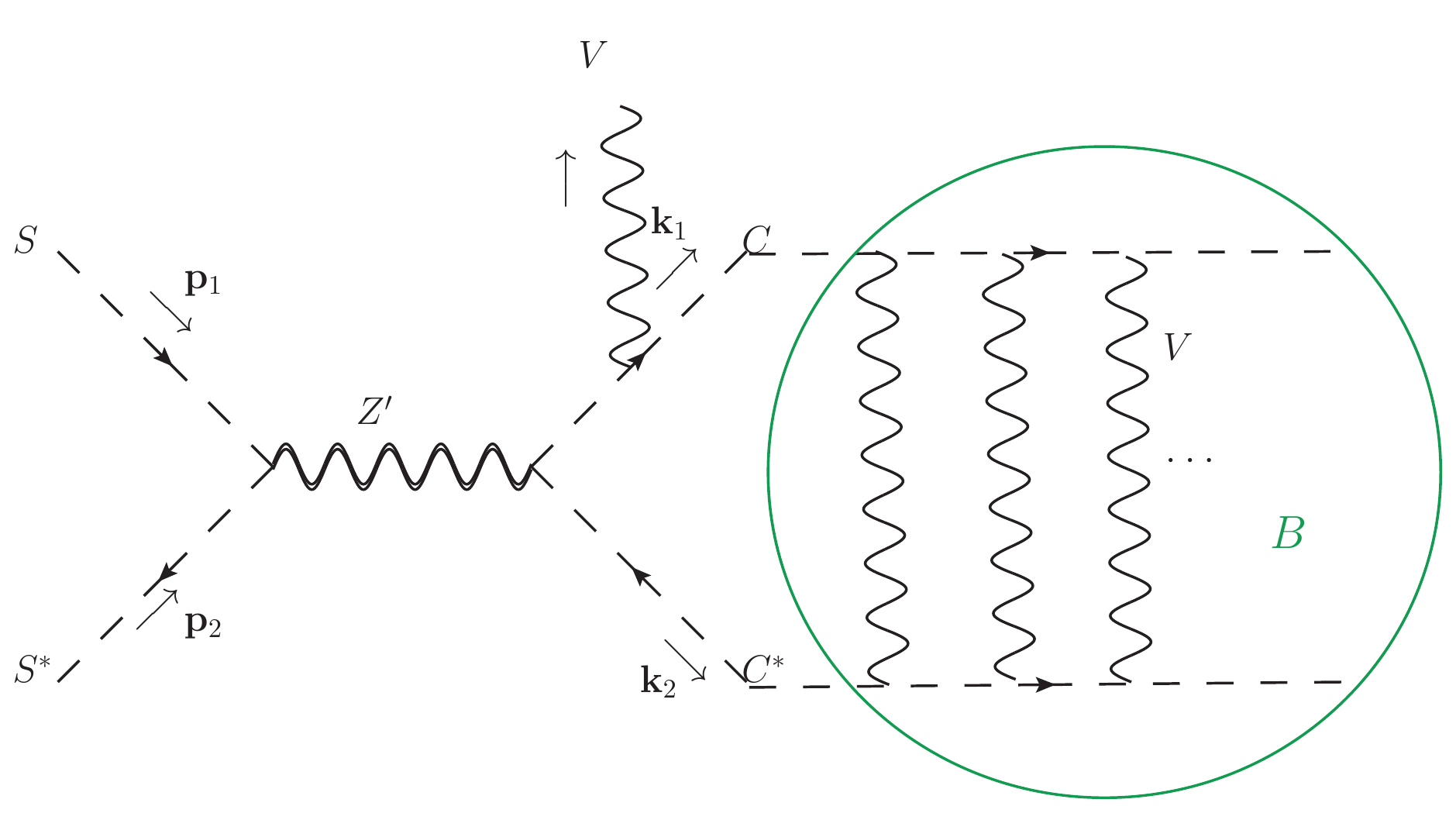}}
  \hspace{0in}
  \subfigure{
  \includegraphics[width=0.4\textwidth]{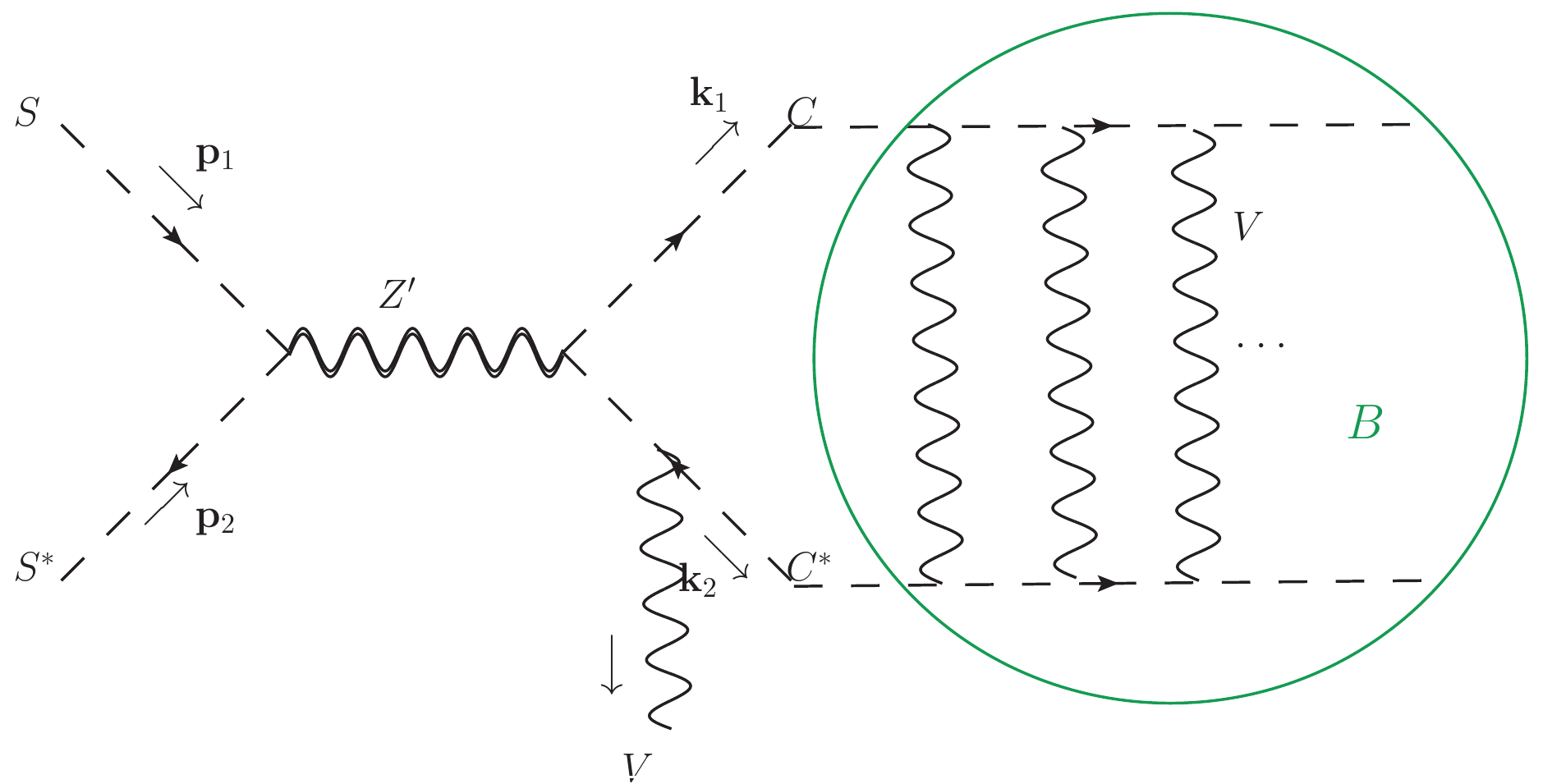}}
   \hspace{0in}
  \subfigure{
  \includegraphics[width=0.4\textwidth]{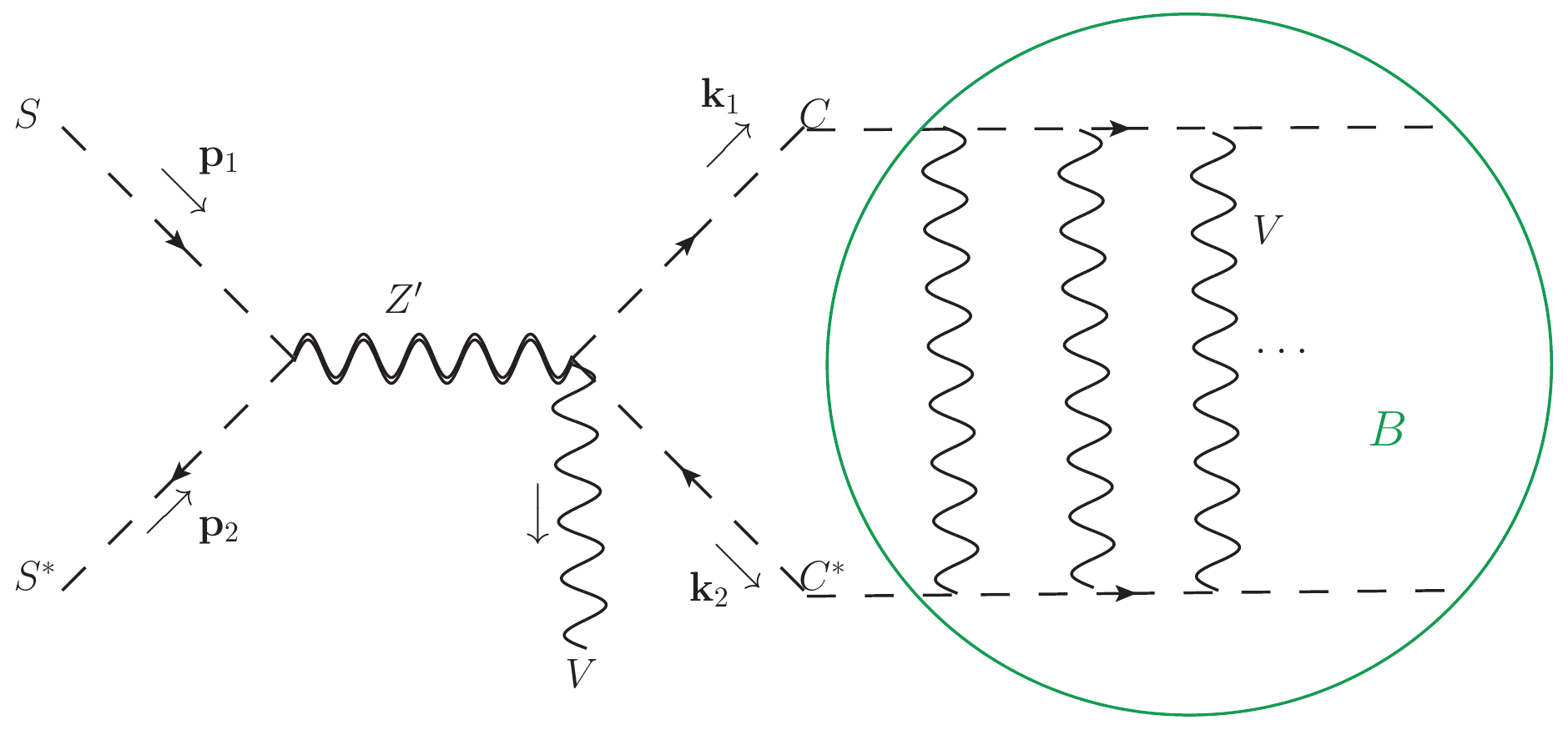}}
  \caption{\label{fig:SSBV} \it The Feynman diagrams for the $(CC^*)$ bound state production process $SS^*\rightarrow BV$.}
\end{figure}

The FBS formation process $SS^*\rightarrow BV$ can be described by three Feynman diagrams shown in Fig.~\ref{fig:SSBV}. The excessive energy can be carried away by a vector boson $V$ emission.

The scattering amplitude for this process can be written as
\begin{equation}
  \label{SSCCV}
  \begin{aligned}
&i\mathcal{M}_{SS^*\rightarrow CC^*V}\\
=&(-ig_6)(p_1-p_2)^\mu\frac{-i}{(p_1+p_2)^2-m_{Z^\prime}^2}\left[g_{\mu\nu}-\frac{(p_1+p_2)_\mu(k_1+k_2+q)_\nu}{m_{Z^\prime}^2}\right]\\
&\times(-ig_5)(k_1+q-k_2)^\nu\frac{i}{(k_1+q)^2-m_C^2}(-ig_3)(k_1+k_1+q)^\beta \cdot\epsilon _\beta^*\\
&+(-ig_6)(p_1-p_2)^\mu\frac{-i}{(p_1+p_2)^2-m_{Z^\prime}^2}\left[g_{\mu\nu}-\frac{(p_1+p_2)_\mu(k_1+k_2+q)_\nu}{m_{Z^\prime}^2}\right]\\
&\times(-ig_5)(k_1-(k_2+q))^\nu\frac{i}{(k_2+q)^2-m_C^2}(-ig_3)(-k_2-(k_2+q))^\beta\cdot\epsilon _\beta^*\\
&+(-ig_6)(p_1-p_2)^\mu\frac{-i}{(p_1+p_2)^2-m_{Z^\prime}^2}\left[g_{\mu\nu}-\frac{(p_1+p_2)_\mu(k_1+k_2+q)_\nu}{m_{Z^\prime}^2}\right]\\
&\times 2ig_3g_5g^{\nu\beta}\cdot\epsilon _\beta^*.
  \end{aligned}
\end{equation}

Because the final state emits a spin one vector boson, considering the angular momentum conservation, the FBS should be $s$-wave. Turning to the rest frame of the ($CC^*$) bound state, we rewrite Eq. \eqref{SSCCV} under non-relativistic approximation and expand to the $zeroth$ order of the final state relative momentum $\mathbf{k}$, as
\begin{equation}
  \mathcal{M}^j_{SS^*\rightarrow CC^*V}=\frac{-2g_3g_5g_6}{s-m_{Z^\prime}^2}\left[(p_1-p_2)^j-\frac{q^j}{2m_C\omega+m_V^2}((p_1-p_2)\cdot q)\right].
\end{equation}
where the index $j$ stands for spatial 3-component.

The FBS effect can be calculated as Model I shows, matrix element multiply the Fourier transform mode of the wave function ($L=0$) which are constituted by $CC^*$ pair relative momentum $\mathbf{k}$, and integrate out the relative momentum $\mathbf{k}$. We get the scattering amplitude for the process $SS^*\rightarrow BV$
\begin{equation}
  \begin{aligned}
  \mathcal{M}^j_{BV} & = \sqrt{\frac{1}{m_C}}\int \frac{d^3\mathbf{k}}{(2\pi)^3}\widetilde{\psi}^*(\mathbf{k}) \ \mathcal{M}^j_{SS^*\rightarrow CC^*V}\\
  & = \sqrt{\frac{1}{m_C}}\psi^*(0) \ \mathcal{M}^j_{SS^*\rightarrow CC^*V}.
  \end{aligned}
\end{equation}
Similarly, the $s$-wave bound state formation scattering amplitude is proportional to the value of wave function at $\mathbf{r}=0$. The Hydrogen-like wave function used here is the $L=0$ part, same as Eq. \eqref{psi}.

Sum over the polarization of the vector boson $V$, then we obtain
\begin{equation}
\begin{aligned}
  &\sum_{\epsilon}|\mathcal{M}_{SS^*\rightarrow BV}|^2\\
  =&(\frac{-2g_3g_5g_6}{s-m_{Z^\prime}^2})^2\frac{1}{m_C}\frac{1}{\pi(na_0)^3}[(s-4m_S ^2)-\frac{(2|\mathbf{p}||\mathbf{q}|\cos\theta)^2}{|\mathbf{q}|^2+m_V^2}\\
  &+\frac{(-2|\mathbf{p}||\mathbf{q}|\cos\theta)^2|\mathbf{q}|^2}{(2m_C\omega+m_V^2)^2}\left(1-\frac{|\mathbf{q}|^2}{|\mathbf{q}|^2+m_V^2}\right)\\
  &+\frac{8(|\mathbf{p}||\mathbf{q}|\cos\theta)^2}{2m_C\omega+m_V^2}\left(1-\frac{|\mathbf{q}|^2}{|\mathbf{q}|^2+m_V^2}\right)].
\end{aligned}
\end{equation}
The cross section times relative velocity $v$ of the incoming DM particles for the process $SS^*\rightarrow B(L=0)V$ is

\begin{equation}
\label{eq:cross section of SS-BV}
  (\sigma v)_{SS^*\rightarrow BV}=C\left[\left(3-B\right)+2A\left(1-B\right)+A^2\left(1-B\right)\right],
\end{equation}
where
\begin{equation}
    \begin{aligned}
      A=&\frac{|\mathbf{q}|_{cm}^2}{2m_C\omega+m_V^2}\\
    B=&\frac{|\mathbf{q}|_{cm}^2}{|\mathbf{q}|_{cm}^2+m_V^2}\\
    C=&\frac{\alpha_{V}^4 g_5^2g_6^2}{6\pi n^3} \frac{m_C^2 (s-4m_S ^2)}{s(s-m_{Z^\prime}^2)^2}\frac{|\mathbf{q}|_{cm}}{\sqrt{s}}.
    \end{aligned}
\end{equation}
In the low temperature limit, $m_V \sim g T = 0$, only the first term in the square brackets is left in Eq. \eqref{eq:cross section of SS-BV}
\begin{equation}
     (\sigma v)_{SS^*\rightarrow BV}=
     \frac{\alpha_{V}^4 g_5^2g_6^2}{3\pi n^3} \frac{m_C^2 (s-4m_S ^2)}{s(s-m_{Z^\prime}^2)^2}\frac{|\mathbf{q}|_{cm}}{\sqrt{s}}.
\end{equation}
This term comes from the third diagram in Fig. \ref{fig:SSBV}, so in Model II, there is no infrared divergence for the $SS^*\rightarrow BV$ process. 
The emitted vector boson energy in the rest frame of the bound state is 
\begin{equation}
  \begin{aligned}
  & \omega =\frac{s-m_B^2-m_V^2}{2m_B}\\
  \end{aligned}
\end{equation}
The formula about $|\mathbf{q}|_{cm}$ is same as what we give in Eq. \eqref{eq:qcm}. It is obvious that the vector boson momentum $|\mathbf{q}|_{cm}$ must be larger than $zero$. It decides the minimum $s_{min}$. 

\subsection{Numerical results of Model II}
\label{sec:Numerical results of Model II}

Like Section \ref{sec:numerical result Model I}, we plot the thermal-averaged cross sections  as a function of the mass ratio of the final and initial state particles at three parameters $\alpha_{V} = 0.02, 0.1, 0.5 $. We normalize other particles mass by DM mass, $z\equiv m_C/m_S$, then we plot the thermal-averaged cross sections evolution as a function of $x$ at three parameters $z = 0.9, 1, 1.1$ and $\alpha_{V} = 0.5 $.

\subsubsection{Thermal averaged cross section}
\label{Thermal averaged cross section II}

The cross sections and kinematic threshold $s_{min}$ for three processes are summarised in Table. \ref{table:cross sections II} In the numerical calculation, we only consider $n=2$ for the $SS^*\rightarrow B$ and $n=1$ for the $SS^*\rightarrow BV$.
\begin{table}[H]
  \centering
  \caption{Cross sections and kinematical forbidden limits for Model II.}
  \begin{tabular}{|l|c|c|c|}
  \hline
  channel & $(\sigma v)$ &  $s_{min}$ \\
  \hline 
      $SS^*\rightarrow CC^*$ & $\frac{g_5 ^2 g_6 ^2}{24 \pi s} \frac{(s-4m_S^2)(s-4m_C ^2)}{(s-m_{Z^\prime}^2)^2} v_2(1+(\frac{\alpha_{V}}{2v_2})^2)\frac{\pi\alpha_{V}/v_2}{1-e^{-\pi\alpha_{V}/v_2}}$ & $Max[4 m_S^2,4m_C^2]$  \\
      $SS^*\rightarrow B$ & $\frac{\alpha_{V} ^5 g_5^2g_6^2 }{12s}\frac{m_C^4 (s-4m_S^2)}{(s-m_{Z^\prime}^2)^2} (\frac{1}{n^3}-\frac{1}{n^5})\delta \left(E_{cm}^2 - m_B^2\right)$ & $4 m_S^2$\\ 
      $SS^*\rightarrow BV$ &$C\left[\left(3-B\right)+2A\left(1-B\right)+A^2\left(1-B\right)\right]$ &$Max[4 m_S^2, \ (m_B+m_V)^2]$ \\
  \hline
  \end{tabular}
  \label{table:cross sections II}
\end{table}

From the above table, it is obvious that a strong enhancement occurs when the mass of propagator $m_{Z^\prime} \approx  2 m_S$, in fact it's the resonance enhancement \cite{Griest:1990kh,Ibe:2008ye,Guo:2009aj}. In this paper, we do not discuss the details about the resonance enhancement and just focus on the parameter region where $m_{Z^\prime} \gg  m_S $ and using approximation $1/(s-m_{Z^\prime}^2)^2 \to 1/m_{Z^\prime}^4$ for the square of $Z^\prime$ propagator to avoid this effect.

\begin{figure}[H]
  \centering
   \subfigure{
  \includegraphics[width=0.3\textwidth]{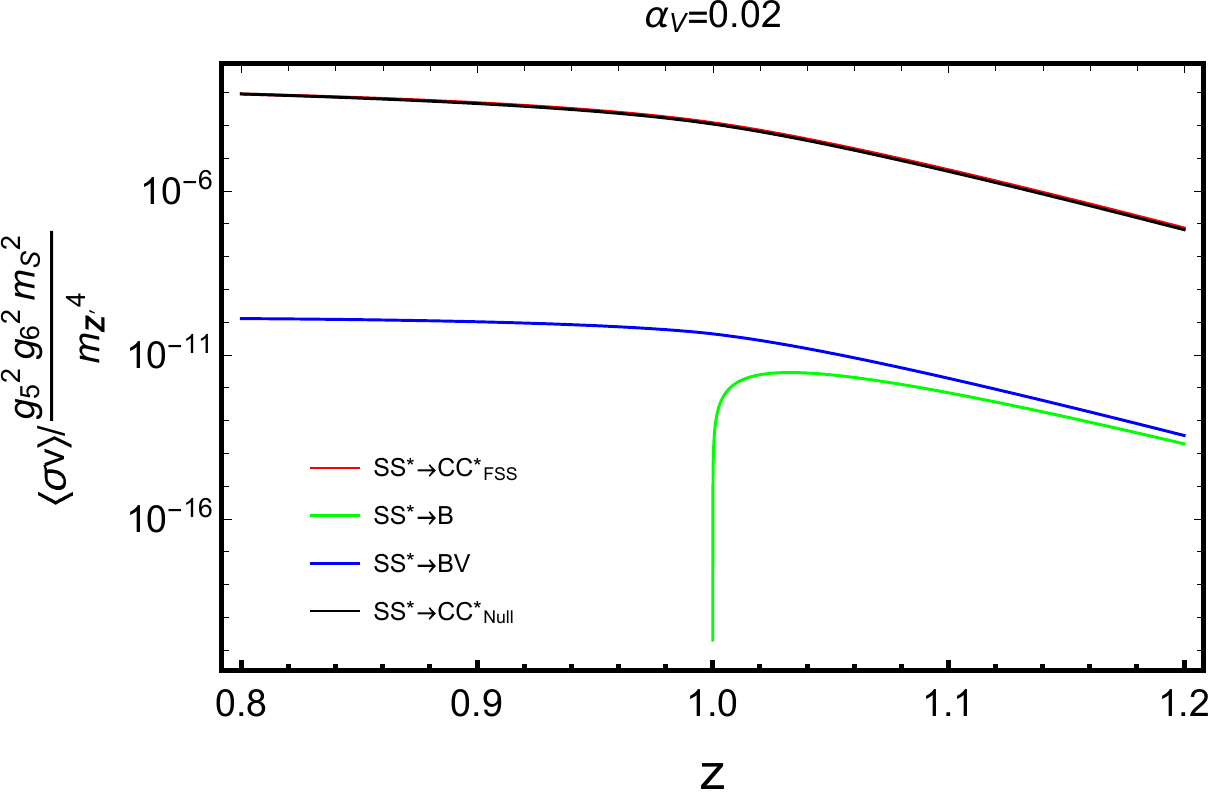}}
  \hspace{0in}
  \subfigure{
  \includegraphics[width=0.3\textwidth]{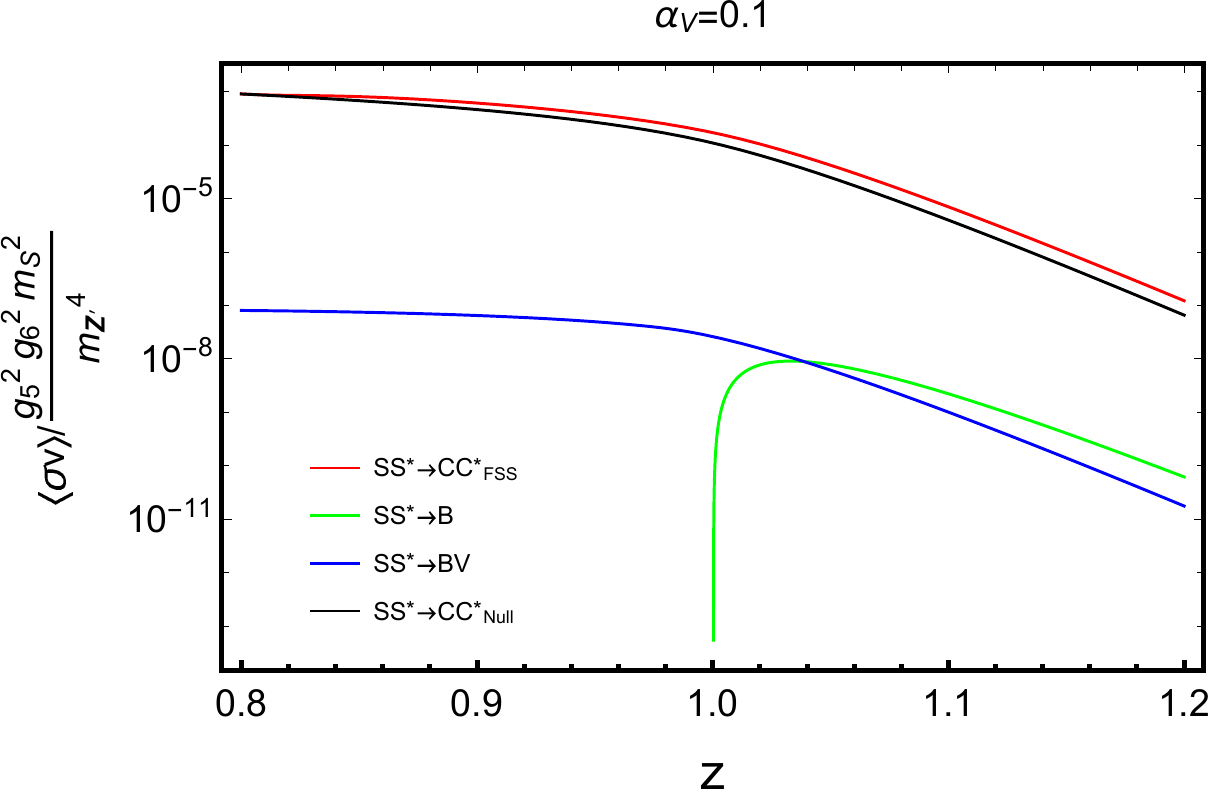}}
  \hspace{0in}
  \subfigure{
  \includegraphics[width=0.3\textwidth]{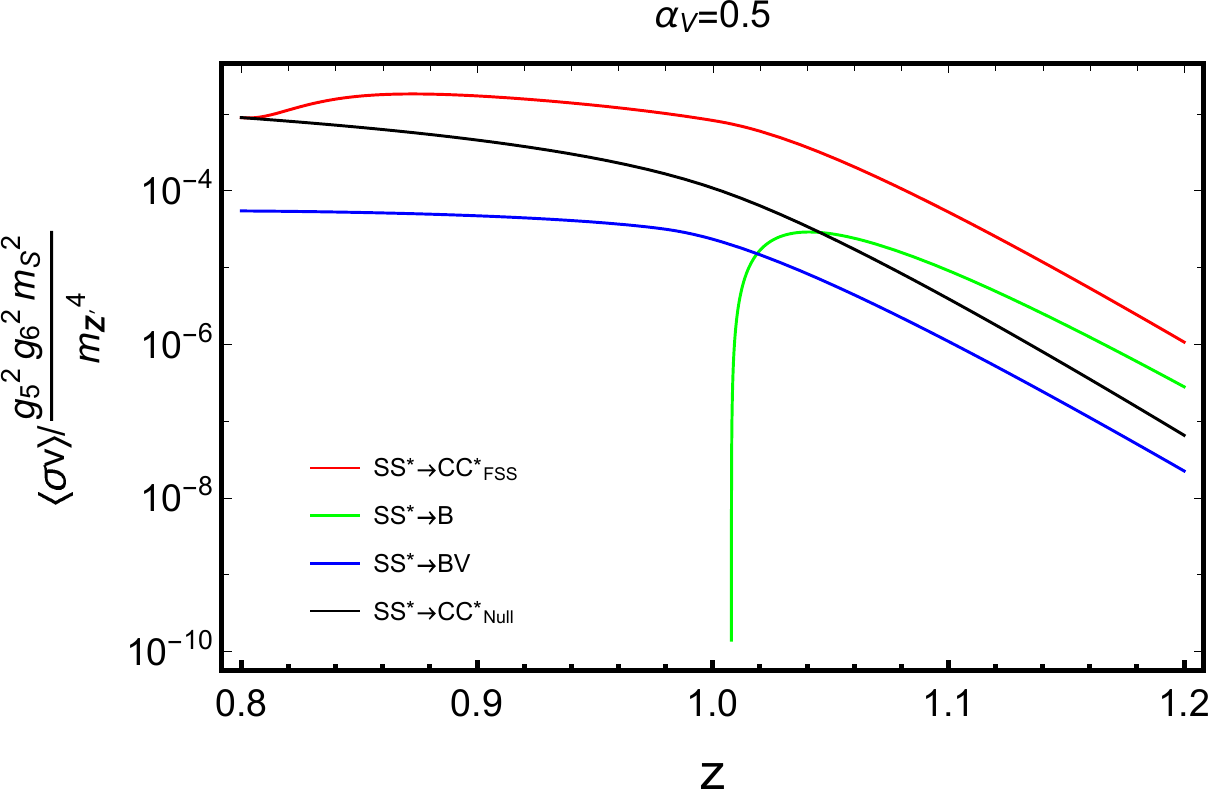}}
  \caption{\label{fig:thermal-averaged cross sections Model II}\it The thermal-averaged cross sections over a common factor $ g_5 ^2 g_6 ^2 m_S ^2/m_{Z^\prime}^4 $ at three parameters $\alpha_{V} = 0.02, 0.1, 0.5$, at a typical freeze-out value $x=m_S/T=25$. The red, green, blue and black lines stand for $\langle\sigma v\rangle_{FSS}$, the thermal-averaged FSS-corrected $p$-wave cross section; $\langle\sigma v \rangle_{B}$, the thermal-averaged FBS (without boson emission) $p$-wave cross section; $\langle\sigma v \rangle_{BV}$, the thermal-averaged FBS (with boson emission) $s$-wave cross section; and $\langle\sigma v \rangle_{w/o both}$, the thermal-averaged cross section without any FSS and FBS, respectively. $z$ is the mass ratio $m_C/m_S$; the $y$-axis is the thermal-averaged cross sections divided by a common factor.}
\end{figure}

We can calculate the thermal-averaged cross sections from Table. \ref{table:cross sections II}, following the Eq. \eqref{average we use} in Section \ref{sec:Thermal averag}. Fig. \ref{fig:thermal-averaged cross sections Model II} shows the thermal-averaged cross sections as a function of the mass ratio of the final and initial state particles at a typical freeze-out value $x=25$. The red, green, blue and black lines stand for $\langle\sigma v\rangle_{FSS}$,  $\langle\sigma v \rangle_{B}$, $\langle\sigma v \rangle_{BV}$ and $\langle\sigma v \rangle_{w/o \ both}$ over a common factor $ g_5 ^2 g_6 ^2 m_S ^2/m_{Z^\prime}^4 $, respectively.

It can be seen from Fig. \ref{fig:thermal-averaged cross sections Model II}, $\langle\sigma v \rangle_{BV}$ can be comparable to $\langle\sigma v \rangle_{B}$, while in model I, the former is much smaller than the latter (Fig. \ref{fig:thermal-averaged cross sections Model I}). The reason is in Model II, it is $s$-wave FBS formation with vector boson emission; while in Model I, it is $p$-wave FBS formation with vector boson emission. In Model II, FBS formation without emission is $p$-wave, its contribution is suppressed comparing to that in Model I in which FBS formation without emission is $s$-wave. On the other hand, at $\alpha_{V}=0.5$, while in model I, $\langle\sigma v \rangle_{B}$ is larger than $\langle\sigma v \rangle_{FSS}$, the former is still smaller than the latter, though they are getting closer.

\begin{figure}[H]
  \centering
   \subfigure{
  \includegraphics[width=0.3\textwidth]{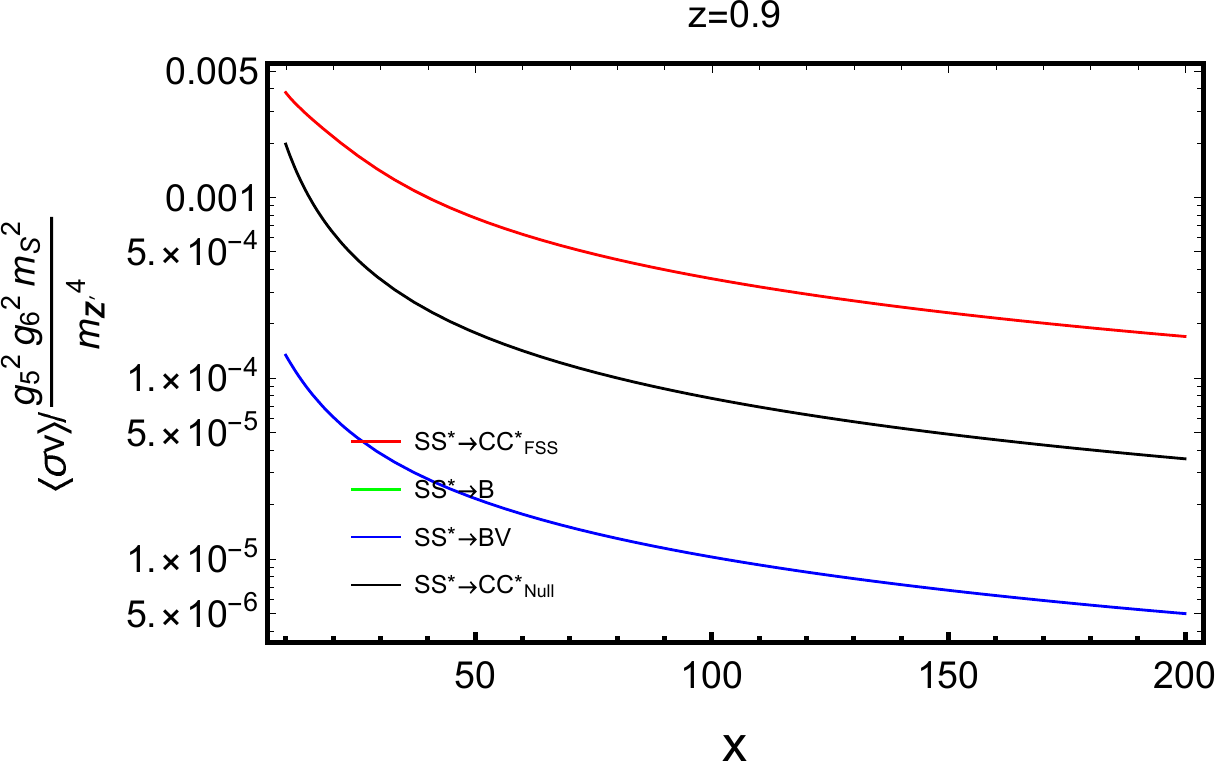}}
  \hspace{0in}
  \subfigure{
  \includegraphics[width=0.3\textwidth]{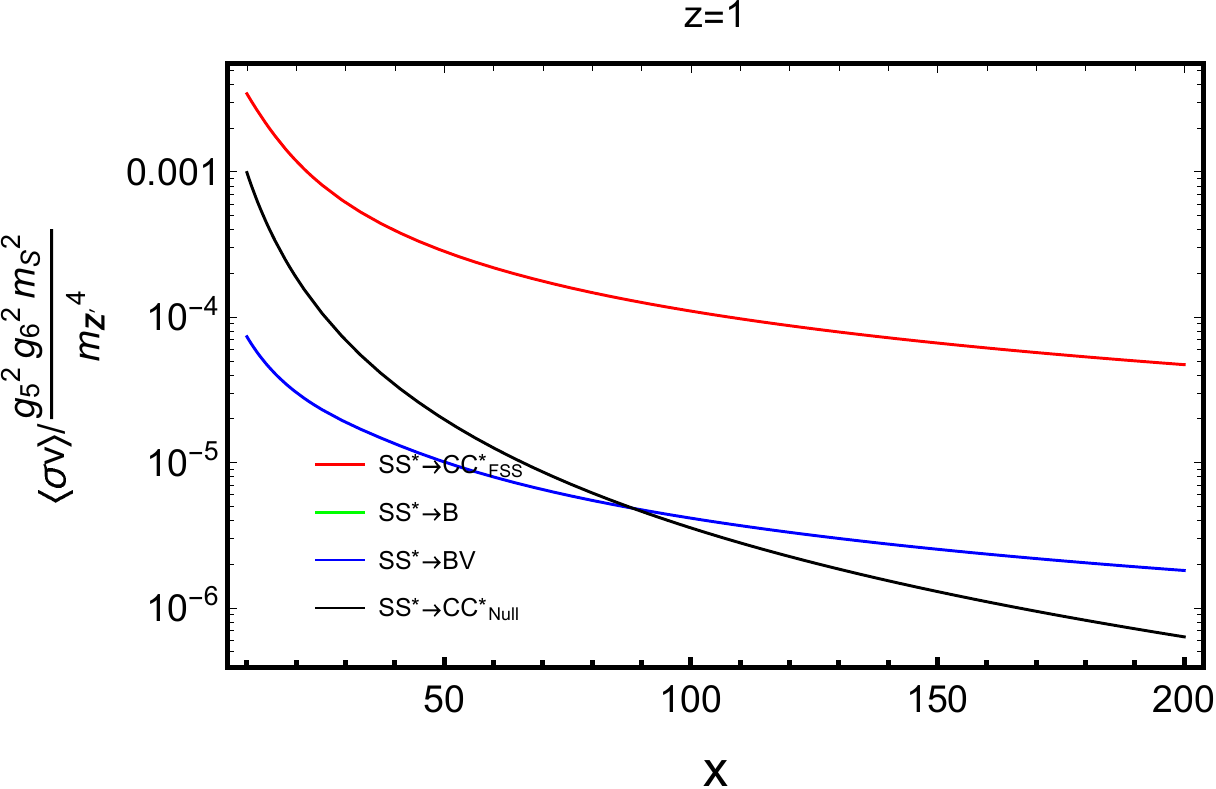}}
  \hspace{0in}
  \subfigure{
  \includegraphics[width=0.3\textwidth]{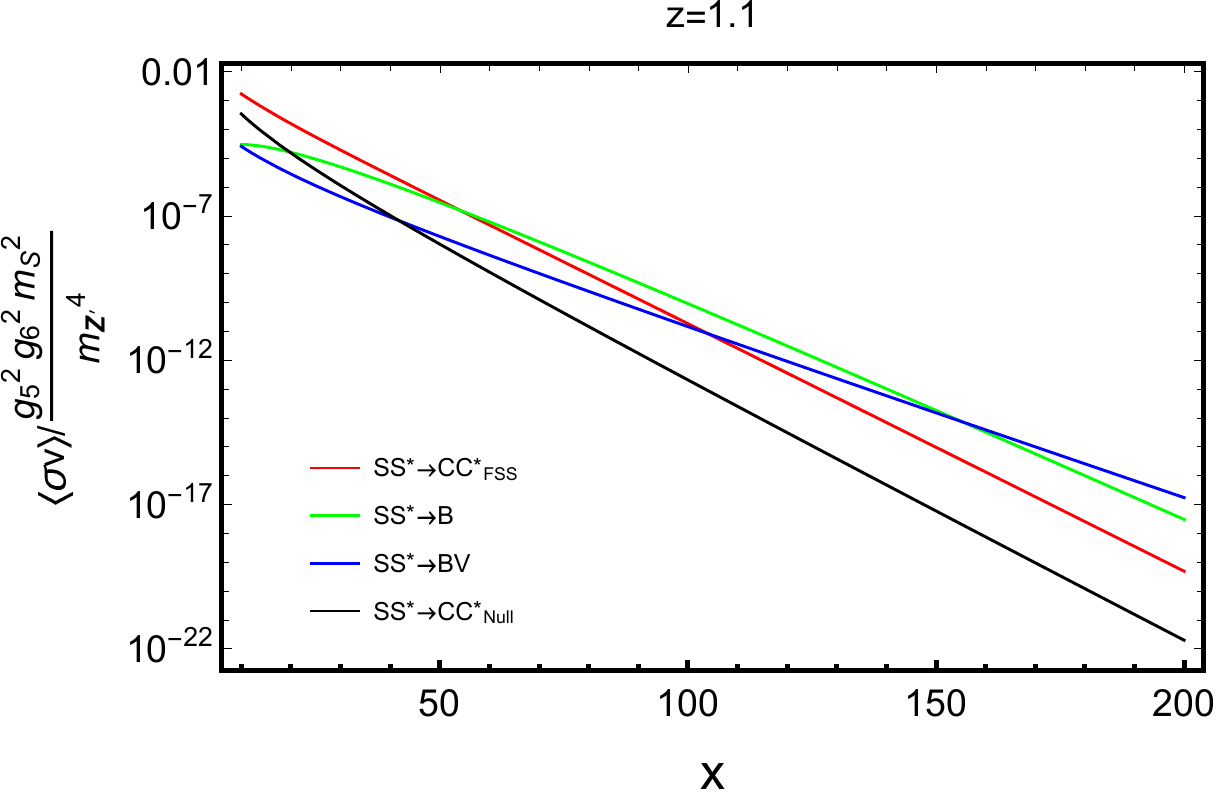}}
  \caption{\label{fig:thermal-averaged cross sections Model II x}\it The thermal-averaged cross sections over a common factor $ g_5 ^2 g_6 ^2 m_S ^2/m_{Z^\prime}^4 $ at three parameters $z = 0.9, 1, 1.1 $ and $\alpha_{V}=0.5$. The red, green, blue and black lines stand for $\langle\sigma v\rangle_{FSS}$, the thermal-averaged FSS-corrected $p$-wave cross section; $\langle\sigma v \rangle_{B}$, the thermal-averaged FBS (without boson emission) $p$-wave cross section; $\langle\sigma v \rangle_{BV}$, the thermal-averaged FBS (with boson emission) $s$-wave cross section; and $\langle\sigma v \rangle_{w/o both}$, the thermal-averaged cross section without any FSS and FBS, respectively. $z$ is the mass ratio $m_C/m_S$; the $y$-axis is the thermal-averaged cross sections divided by a common factor.}
\end{figure}

Fig. \ref{fig:thermal-averaged cross sections Model II x} shows the evolution of thermal-averaged cross sections of Model II as temperature drops. In the left and the middle panels, the $\langle\sigma v \rangle_{B}$ does not appear for the same reason as model I. The right panel is for the forbidden case. It is the same as model I, at low temperature, fewer initial particles can reach the threshold, so the thermal-averaged cross sections become smaller as temperature decreases. However, in the left and the middle panel, the trend of the lines are different in Fig.\ref{fig:thermal-averaged cross sections Model I,f(x)} and Fig.\ref{fig:thermal-averaged cross sections Model II x}. It is because in Model II, the three point vertex are proportional to the momentum, that is why the thermal-averaged cross sections are smaller with the temperature decreasing.

The right panel of Fig. \ref{fig:thermal-averaged cross sections Model II x} indicates that the $SS^*\rightarrow BV$ prosses may provide a significant indirect detection signal for forbidden dark matter. Because the process $SS^\ast \to BV$ becomes more and more important as $x$ decreases, and it dominates in the low temperature ($x>150$).

\subsubsection{Relic abundance}

\begin{figure}[H]
  \centering
   \subfigure{
  \includegraphics[width=0.3\textwidth]{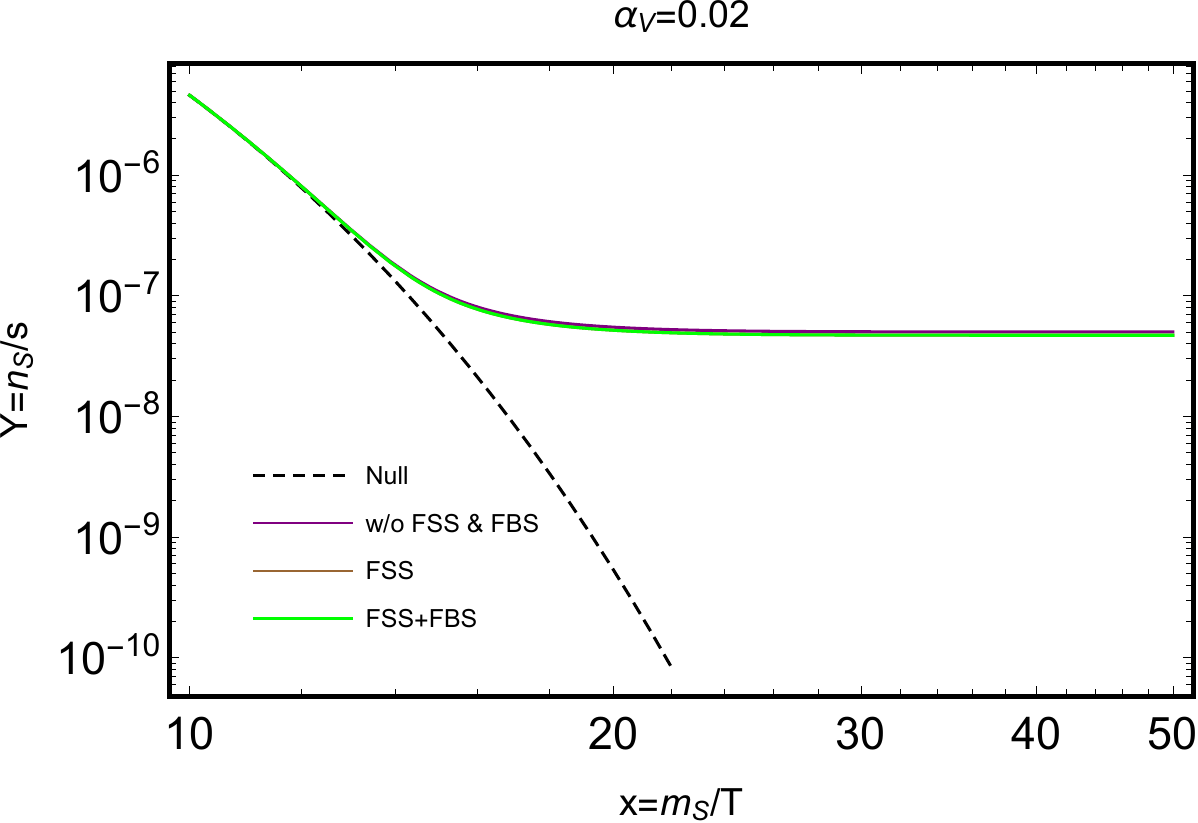}}
  \hspace{0in}
  \subfigure{
  \includegraphics[width=0.3\textwidth]{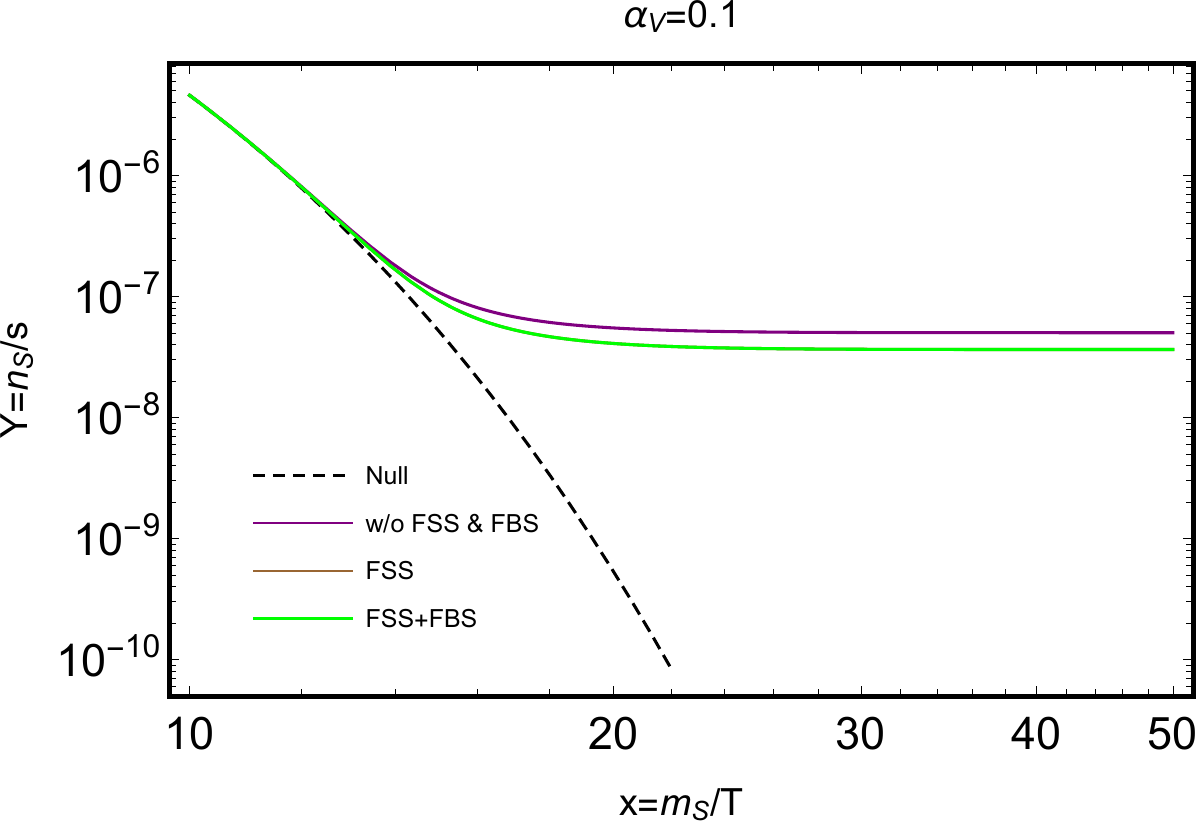}}
  \hspace{0in}
  \subfigure{
  \includegraphics[width=0.3\textwidth]{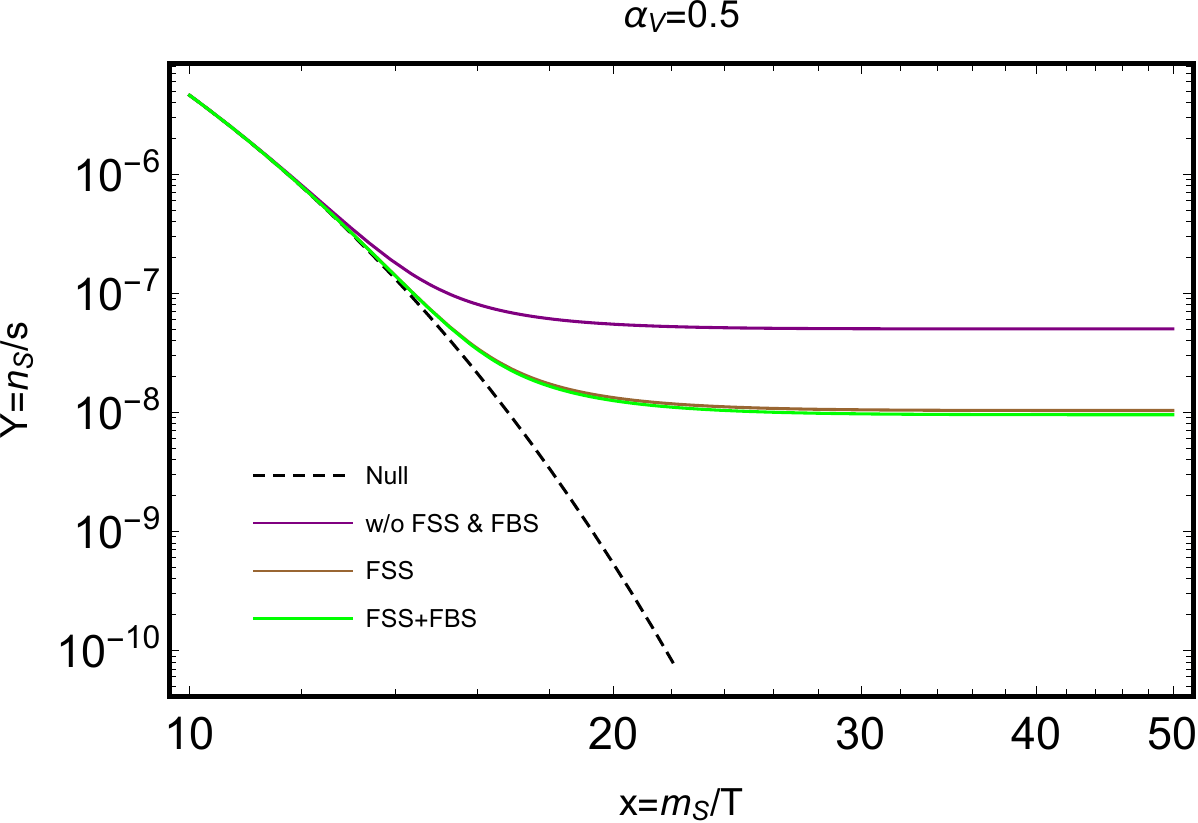}}
  \caption{\label{fig:model_II_abundance1}\it The evolution of the DM $yield$ as a function of $x=m_S/T$ for the representative case $m_S=500 GeV$, $z=1.1$, $ g_5 ^2 g_6 ^2 m_S ^2/m_{Z^\prime}^4= 10^{-6} GeV^{-2}$ and $\alpha_{V}=0.02,0.1,0.5$. The purple line neglects both FSS and FBS effects. The brown line neglects FBS effect. The green line incorporates the effects of FBS and FSS. The black dashed line exhibits the naive thermal equilibrium abundance. }
\end{figure}

We have already obtained the thermal-averaged cross sections numerically in \ref{Thermal averaged cross section II}, it is not hard to solve the Boltzmann equation as Section \ref{sec:Boltzmann equation} outlined by substituting the $m_D$ to $m_S$. Again, we choose three parameters $\alpha_{V} = 0.02, 0.1, 0.5 $, and show the $yield$ of DM as a function of $x$ considering different effects. Other parameters are chosen as $z=1.1$ (the forbidden case), $m_S=500 GeV$, $ g_5 ^2 g_6 ^2 m_S ^2/m_{Z^\prime}^4= 10^{-6} GeV^{-2}$. As Fig. \ref{fig:model_II_abundance1} shows, the purple line neglects both FSS and FBS effects, the brown line neglects FBS effect, the green line incorporates the effects of FBS and FSS.

Compared to Model I, the FBS effect is milder, the $p$-wave FSS effect also has a significant enhancement on DM annihilation.  Because in Model II, the FBS effect without emission is $p$-wave. Although the FBS effect with emission is $s$-wave, it is still suppressed by order $\alpha_{V}$ because a vector boson emission. But in the right panel, $\alpha_{V}=0.5$, the FBS effect contribution is still visible, it further reduces the relic abundance by $13\%$ on top of the FSS effect.

Fig. \ref{fig:thermal-averaged cross sections Model II} and \ref{fig:model_II_abundance1} show that the FBS formation effect and FSS effects are important in DM relic abundance calculation when DM annihilation products are non-relativistic and have a large coupling with a light vector boson. 

\section{Conclusion}
\label{sec:comclusion}




We have studied the FBS effect on DM relic abundance in this paper. We employ two DM models, and calculate the FSS effect and FBS effect using Coulomb-like potential approximation. We give the numerical results considering those effects, which demonstrate that the FBS has a significant effect on the DM relic abundance if DM annihilation products move non-relativistically and there is some long-range force between them, particularly in the forbidden dark matter cases. 

Compared to previous works on this subject, the FBS effect which had not been previously taken into account expands the scope of both DM abundance calculation and the complementary ways of experimental detection. We point out the following salient features of this work: 

$a$. Most of the previous work focus on the initial state bound state (IBS) effect, We stress that the same argument could be extended to FBS effect. We provide two models to show that the FBS effect has a significant influence on the DM relic abundance compared to the FSS effect, especially for the ``forbidden'' cases. 

$b$. We find the usual sub-leading FBS formation process can not be negligible compared to the lead process, and has the potential to give the indirect detect signals. 

$c$. We also consider the $p$-wave FSS effect in this work comparing to \cite{Cui:2020ppc}. 



\section*{Acknowledgements}
F.L. thanks Xiaoyi Cui and Shu Lin for helpful discussions. 
The work of F.L. is supported in part by the One Hundred Talent Program of Sun Yat-sen University, China.

\bibliography{ref}
\bibliographystyle{unsrt}

\end{document}